\documentclass[pdflatex,sn-mathphys-num]{sn-jnl}
\usepackage[switch]{lineno}  

\usepackage{graphicx}%
\usepackage{multirow}%
\usepackage{amsmath,amssymb,amsfonts}%
\usepackage{amsthm}%
\usepackage{mathrsfs}%
\usepackage[title]{appendix}%
\usepackage{xcolor}%
\usepackage{textcomp}%
\usepackage{manyfoot}%
\usepackage{booktabs}%
\usepackage{algorithm}%
\usepackage{algorithmicx}%
\usepackage{algpseudocode}%
\usepackage{listings}%

\theoremstyle{thmstyleone}%

\theoremstyle{thmstyletwo}%

\theoremstyle{thmstylethree}%

\begin{document}

\title[Article Title]{Wavelet-SARIMA-Transformer: A Hybrid Model for Rainfall Forecasting}

\author[1]{\fnm{Junmoni} \sur{Saikia}}\email{junmonisaikia1302@gmail.com}

\author*[1,2]{\fnm{Kuldeep} \sur{Goswami}}\email{kgkuldeepgoswami@gmail.com}

\author[1]{\fnm{Sarat C.} \sur{Kakaty}}\email{saratkakaty@gmail.com}

\affil*[1]{\orgdiv{Department of Statistics}, \orgname{Dibrugarh University}, \orgaddress{\city{Dibrugarh}, \postcode{786004}, \state{Assam}, \country{India}}}

\affil[2]{\orgname{Srushti Drushti Geospatial Foundation}, \orgaddress{\city{Rampurhat}, \state{West Bangal}, \country{India}}}


\abstract{
Rainfall forecasting in monsoon-dominated regions is challenging due to the nonlinear, nonstationary, and scale-dependent nature of precipitation dynamics. This study develops and evaluates a novel hybrid Wavelet–SARIMA–Transformer (W-ST) framework to forecast using monthly rainfall across five meteorological subdivisions of Northeast India over the 1971–2023 period. The approach employs the Maximal Overlap Discrete Wavelet Transform (MODWT) with four wavelet families (Haar, Daubechies, Symlet, Coiflet) to achieve shift-invariant, multiresolution decomposition of the rainfall series. Linear and seasonal components are modeled using Seasonal ARIMA (SARIMA), while nonlinear components are modeled by a Transformer network, and forecasts are reconstructed via inverse MODWT.  

Comprehensive validation using an 80:20 train–test split and multiple performance indices (RMSE, MAE, SMAPE, Willmott’s $d$, Skill Score, Percent Bias, Explained Variance, and Legates–McCabe’s $E_1$) demonstrates the superiority of the Haar-based hybrid model [W(H)-ST]. Across all subdivisions, W(H)-ST consistently achieved lower forecast errors, stronger agreement with observed rainfall, and unbiased predictions compared with stand-alone SARIMA, stand-alone Transformer, and two-stage wavelet hybrids. Residual adequacy was confirmed through the Ljung–Box test, while Taylor diagrams provided an integrated assessment of correlation, variance fidelity, and RMSE, further reinforcing the robustness of the proposed approach.  

The results highlight the effectiveness of integrating multiresolution signal decomposition with complementary linear and deep learning models for hydroclimatic forecasting. Beyond rainfall, the proposed W-ST framework offers a scalable methodology for forecasting complex environmental time series, with direct implications for flood risk management, water resources planning, and climate adaptation strategies in data-sparse and climate-sensitive regions.
}




\maketitle

\section{Introduction}\label{sec1}

Time series modeling has long relied on statistical approaches, with the Autoregressive Integrated Moving Average (ARIMA) methodology, formalized by Box and Jenkins \cite{bib29, bib30}, dominating hydrological and meteorological forecasting for decades. Despite its widespread use, ARIMA assumes linearity and stationarity, which limits its ability to capture the complex, nonlinear, and nonstationary nature of rainfall and hydrological time series. Consequently, researchers have increasingly turned to nonparametric and data-driven approaches to overcome these limitations.

Among these, wavelet analysis has emerged as a powerful methodology for decomposing complex signals into multiple scales of time–frequency components. By enabling localized, multiscale feature extraction, wavelet decomposition facilitates improved representation of rainfall dynamics and has been shown to enhance predictive performance when coupled with machine learning (ML) and deep learning (DL) models \cite{bib41, bib42, bib43, bib44, bib45, bib46}. Numerous studies confirm that wavelet–ML/DL hybrids outperform classical models by capturing both linear and nonlinear dependencies inherent in rainfall datasets \cite{bib2, bib5, bib6, bib7, bib11}.

Hybrid models that combine wavelet transforms with traditional approaches (e.g., ARIMA, ANN, LSSVM) have consistently demonstrated superior forecasting accuracy for precipitation, streamflow, and drought prediction across diverse climatic regions \cite{bib6, bib7}. More advanced integrations, such as wavelet–CNN, wavelet–LSTM, and wavelet–neuro-fuzzy systems, have further improved modeling capability by capturing both short-term fluctuations and long-range dependencies in rainfall data \cite{bib8, bib20}. These efforts underscore the importance of multi-resolution decomposition in enhancing model robustness and accuracy.

Parallel to these developments, deep learning architectures have revolutionized hydrometeorological forecasting. Convolutional and recurrent networks, including CNNs, LSTMs, and ConvLSTMs, have demonstrated effectiveness in extracting spatiotemporal rainfall features from large-scale datasets \cite{bib18, bib21, bib4, bib22}. Yet, recurrent models still suffer from gradient vanishing and limited capability in modeling long-range temporal dependencies. In response, transformer architectures, originally designed for natural language processing \cite{bib22, bib23, bib24}, have been successfully adapted to time series prediction \cite{bib26, bib33, bib34}. Transformers leverage self-attention mechanisms to efficiently model long-term dependencies and have been shown to outperform RNNs, GRUs, and LSTMs in several rainfall and hydrological applications \cite{bib28}.

Recent studies highlight the potential of wavelet–transformer hybrids, which combine the strengths of wavelet decomposition (handling nonstationarity and noise) with transformer-based sequence modeling (capturing long-range dependencies). Such models have achieved state-of-the-art performance in diverse domains including rainfall, wind speed forecasting, and even image super-resolution \cite{bib12, bib13, bib14, bib16, bib17, bib27}. These developments reflect an evolutionary trajectory from linear ARIMA models → wavelet-enhanced statistical/ML hybrids → advanced deep learning architectures → wavelet–transformer frameworks, highlighting a growing emphasis on robustness, scalability, and accuracy in hydrometeorological forecasting.

Motivated by these advances, this study proposes a novel Wavelet–SARIMA–Transformer (W-ST) framework for rainfall forecasting. In our approach, wavelet decomposition (using the Maximal Overlap Discrete Wavelet Transform, MODWT) is first employed to separate rainfall series into linear and nonlinear components. The linear sub-series are modeled using SARIMA, while the nonlinear sub-series are predicted using a transformer network, enabling simultaneous exploitation of multi-resolution decomposition and long-range attention mechanisms. By employing diverse wavelet families (Haar, Daubechies, Symlet, Coiflet) and applying the model across multiple regional rainfall datasets, the proposed framework aims to achieve robust, scalable, and accurate rainfall forecasting.

\section{Methodology}\label{sec2}
\begin{figure}[H]
    \centering
    \includegraphics[scale=0.9]{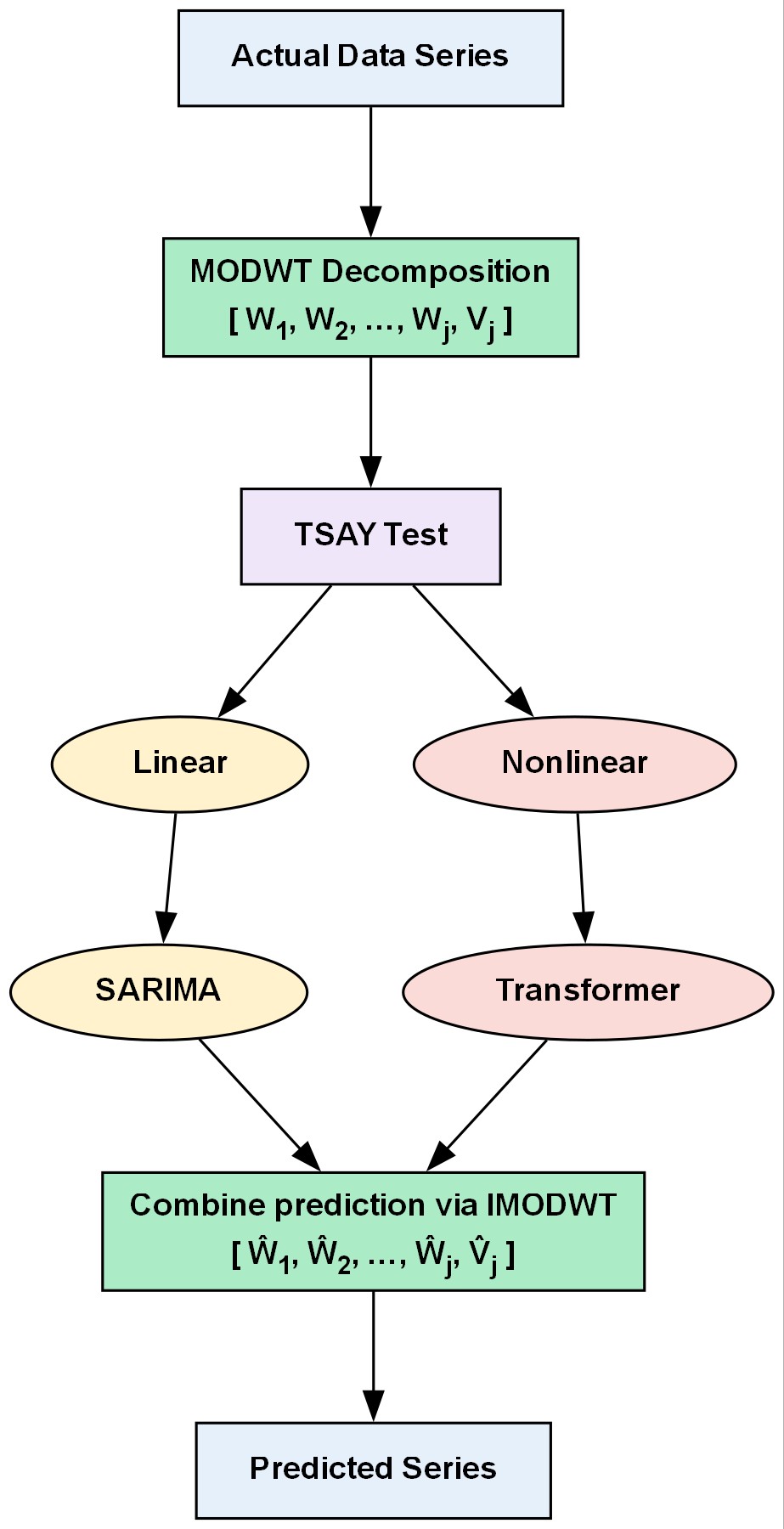}
    \caption{Schematic representation of the proposed Wavelet-SARIMA-Transformer model}
    \label{fig:workflow}
\end{figure}

This study proposes a hybrid \textbf{Wavelet--SARIMA--Transformer (WST)} framework for rainfall forecasting, integrating statistical and deep learning paradigms under a multiresolution wavelet decomposition scheme. The methodology is structured into three components: (i) wavelet-based multiresolution decomposition, (ii) SARIMA modeling for linear patterns, and (iii) Transformer modeling for nonlinear patterns. The workflow culminates in reconstruction using the inverse MODWT, producing a robust hybrid forecast. The schematic diagram of the proposed approach is shown in Fig.~\ref{fig:workflow}.
\subsection{Wavelet Decomposition}\label{subsec2}
Wavelet analysis has become a widely adopted technique for preprocessing signals, particularly in time series analysis. Unlike the traditional Fourier transform, which provides only frequency-domain information, wavelet methods simultaneously capture both temporal and frequency characteristics, making them especially suitable for analyzing nonstationary processes. Among the different approaches, the Discrete Wavelet Transform (DWT) and the Continuous Wavelet Transform (CWT) are the most common. The DWT decomposes a time series into low- and high-frequency components using a predefined filter; however, it suffers from certain limitations such as the requirement that the series length be a power of two ($N=2^p$), downsampling at each level, and translation sensitivity. To overcome these issues, the Maximal Overlap Discrete Wavelet Transform (MODWT) has been introduced \cite{bib50},\cite{bib51}. Unlike the DWT, the MODWT employs filter banks with longer impulse responses and overlapping sub-bands, enabling a more complete and accurate decomposition. Importantly, subsampling is postponed until the final stage, ensuring that the full dataset is retained throughout the decomposition. As a result, the MODWT produces sub-bands of equal length to the original series, preserves alignment with the time index, and facilitates improved signal reconstruction compared to the DWT.\\ 

The original rainfall series $\{X_n\}$ can be decomposed into approximation coefficients $V_{j,n}$ and detail coefficients $W_{j,n}$ at multiple levels $j$ using dilated wavelet filters. The decomposition at each level is given by,
\begin{align}
W_{j, n} &= \sum_{l=0}^{L-1} \tilde{h}_{j, l} \, V_{j-1, (n - 2^{j-1} l) \bmod N}, \label{eq:modwt_detail} \\
V_{j, n} &= \sum_{l=0}^{L-1} \tilde{g}_{j, l} \, V_{j-1, (n - 2^{j-1} l) \bmod N}, \label{eq:modwt_approx}
\end{align}
where,
     $\tilde{h}_{j,l}$ and $\tilde{g}_{j,l}$ are the rescaled wavelet and scaling filters at level $j$, computed as the discrete wavelet filter coefficients divided by $\sqrt{2}$ . $L$ is the filter length and the modulo ensures circular (periodic) convolution and 
    $V_{0,n} = X_n$ is the original rainfall series.

The approximation coefficients on the previous scale $V_{j-1,n}$ can be reconstructed from the detail and the approximation coefficients on scale $j$ using the synthesis filters $\tilde{h}^{\circ}_{j, l}$ and $\tilde{g}^{\circ}_{j, l}$:
\begin{equation}
V_{j-1, n} = \sum_{l=0}^{L-1} \tilde{h}^{\circ}_{j, l} \, W_{j, (n + 2^{j-1} l) \bmod N} + \sum_{l=0}^{L-1} \tilde{g}^{\circ}_{j, l} \, V_{j, (n + 2^{j-1} l) \bmod N}, \label{eq:modwt_recon}
\end{equation}

MODWT-based Multiresolution Analysis (MRA) decomposes the original signal into additive components representing variations at each scale. The detail contribution at scale $j$ is given by,
\begin{equation}
D_{j, n} = \sum_{t=0}^{k-1} \tilde{h}^{\circ}_{j, t} \, W_{j, (n + t) \bmod k}, \label{eq:mra_detail}
\end{equation}
and the smooth (approximation) component at the final scale $J$ is
\begin{equation}
S_{J, n} = \sum_{t=0}^{k-1} \tilde{g}^{\circ}_{J, t} \, V_{J, (n + t) \bmod k}, \label{eq:mra_smooth}
\end{equation}
such that the original time series can be exactly reconstructed as
\begin{equation}
X_n = \sum_{j=1}^{J} D_{j, n} + S_{J, n}, \label{eq:mra_sum}
\end{equation}
where,
  $D_{j,n}$ captures fluctuations at scale $j$ and
$S_{J,n}$ represents the long-term trend after $J$ levels of decomposition.
   
An appropriate selection of a wavelet function is extremely crucial. It is based on attributes such as compact support, smoothness, symmetry, and the number of vanishing moments. The following section outlines the important characteristics of  wavelet families that we are going to use —Haar, Daubechies, Symlets, and Coiflets.

\begin{figure}[htbp]
\centering
    \includegraphics[width=0.45\textwidth]{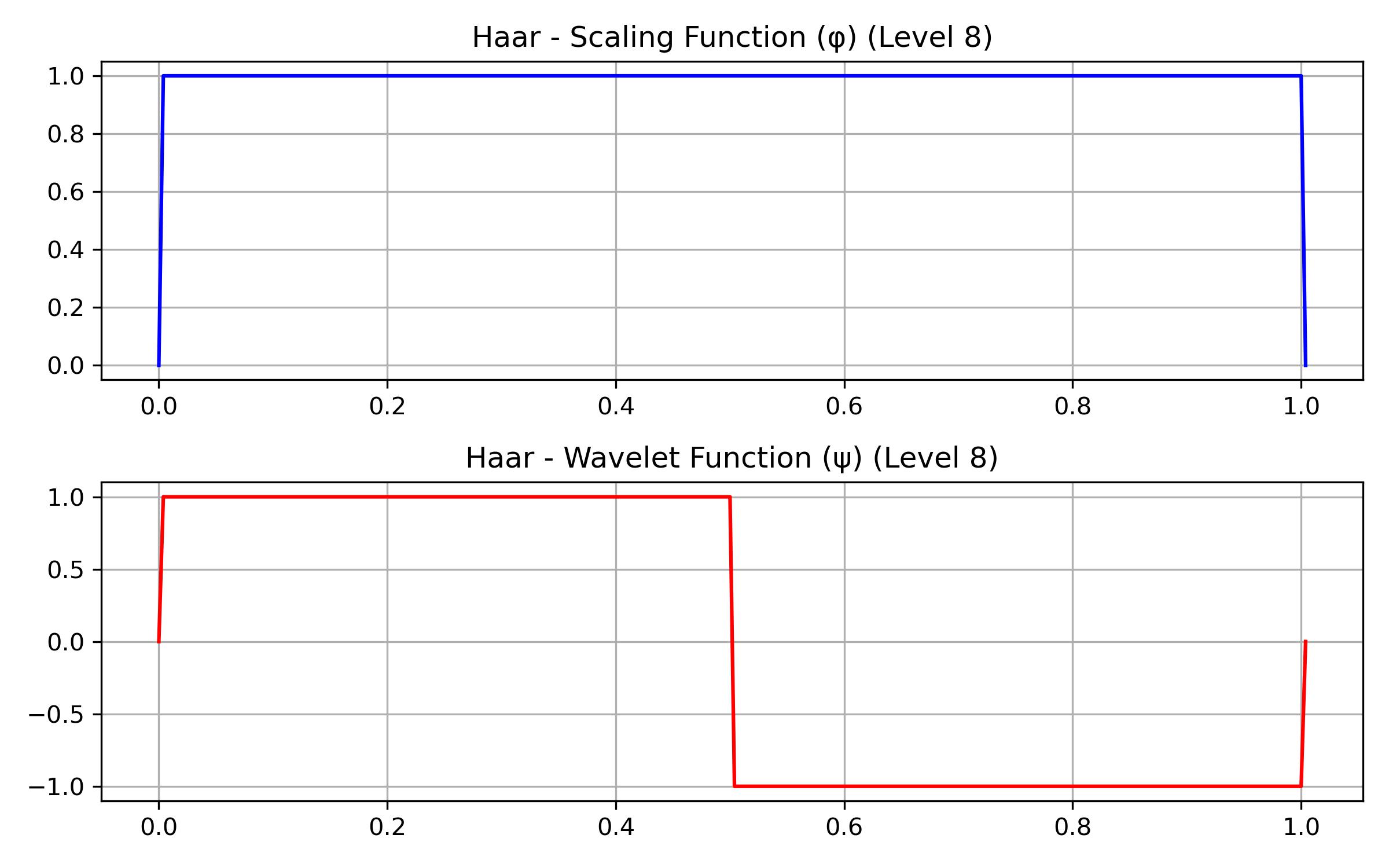}
    \includegraphics[width=0.45\textwidth]{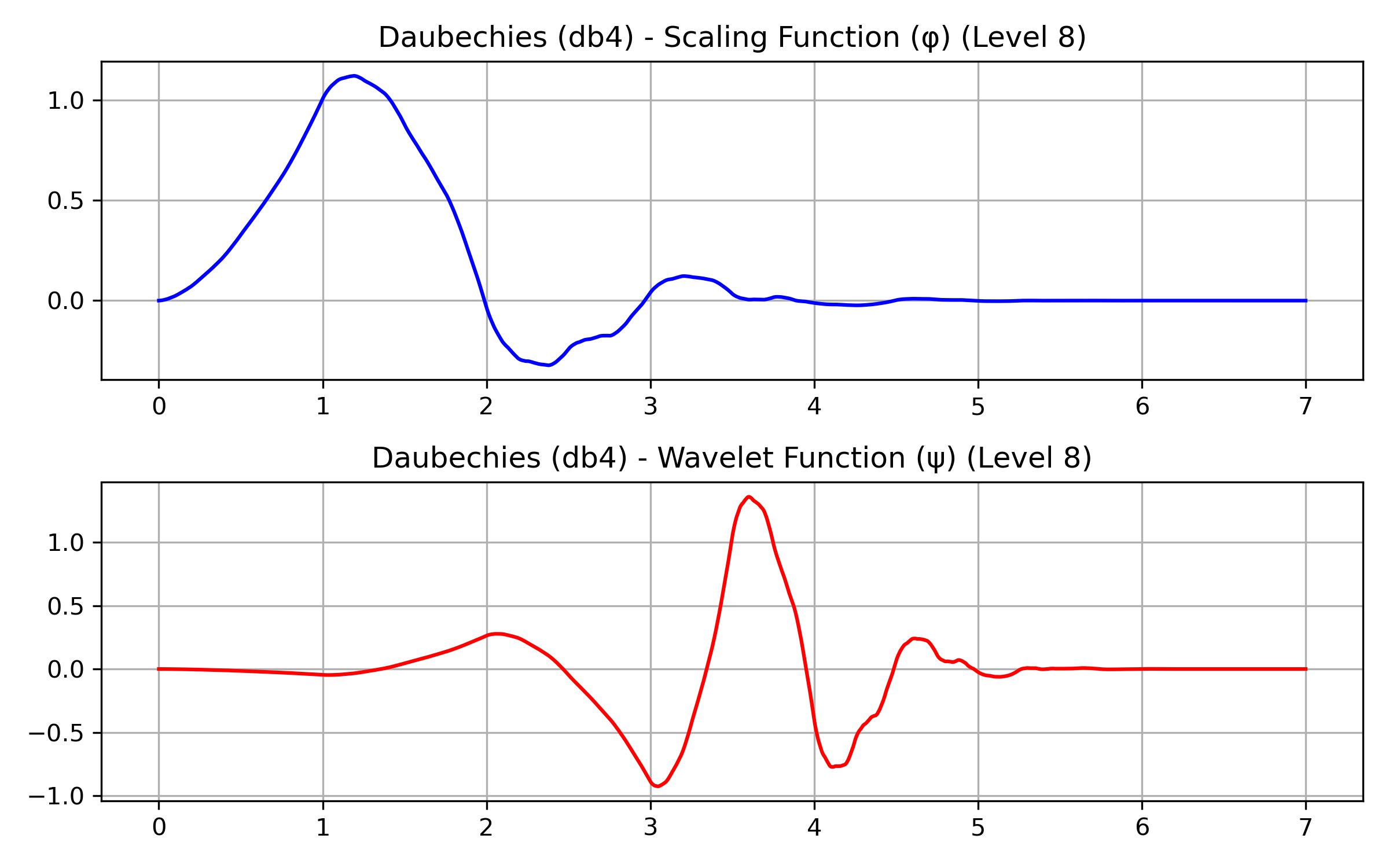}
    \includegraphics[width=0.45\textwidth]{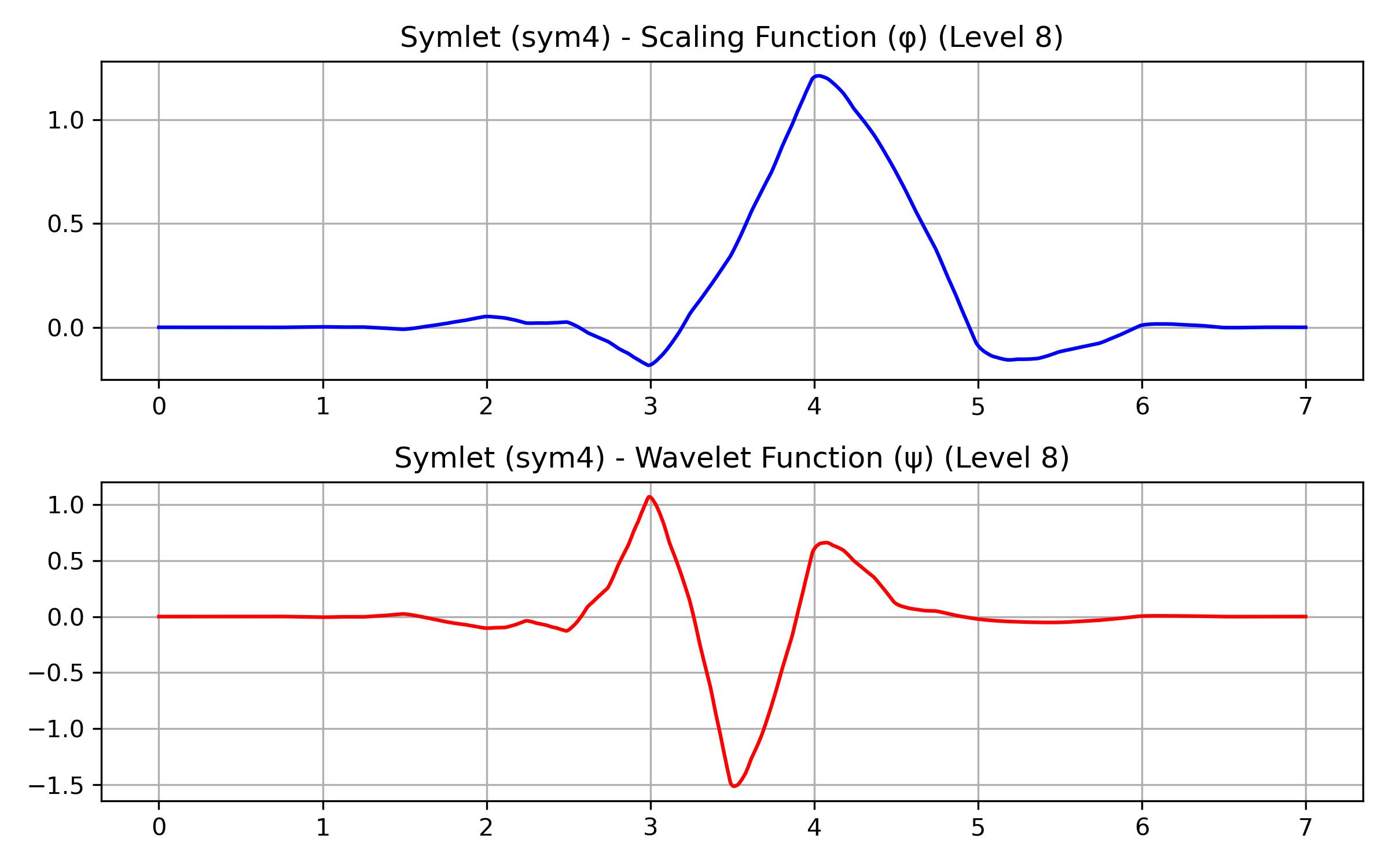}
    \includegraphics[width=0.45\textwidth]{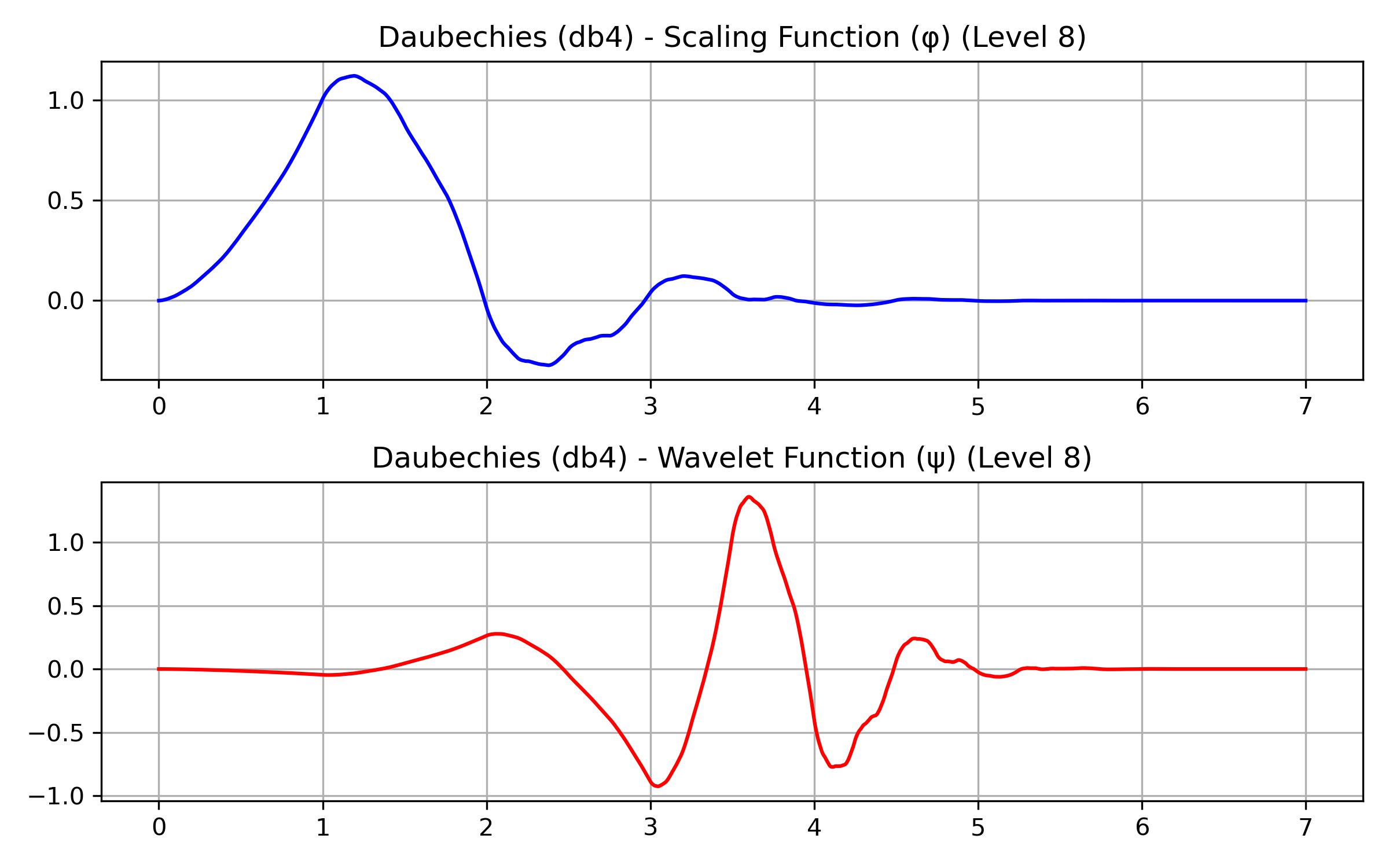}
    \caption{Shape of (a)Haar, (b)Daubechies, (c)Symlet and (d)Coiflet wavelet function}
    \label{fig:wavelets}
\end{figure}

\begin{enumerate}
 \item Haar wavelet- It is named after Alfréd Haar \cite{bib36}, the inventor of the wavelet in 1910, is the oldest orthogonal wavelet. It has one vanishing moment, and it is a discrete first-order difference operator. The piecewise constant nature of the Haar wavelet is what makes it suitable for the detection of discontinuities and sudden changes in time series signals, thereby making it statistically effective in the modeling of non-smooth signals.Fig.~\ref{fig:wavelets}
 
\item Daubechies Wavelet- It is introduced by Ingrid Daubechies in 1992 \cite{bib37}, and are compactly supported orthogonal wavelets characterized by the largest number of vanishing moments for a support width. They are statistically stable to model smooth signals, thus reducing regression estimation error and reducing noise through thresholding in wavelet shrinkage algorithms. Although they can cause boundary distortions because of non-symmetry, they remain an integral part of nonparametric curve estimation and multiscale modeling.Fig.~\ref{fig:wavelets}

\item Symlet Wavelet- These are variants of the original Daubechies wavelets with greater symmetry \cite{bib38}.These have orthogonality and vanishing moment characteristics and reduce phase distortion. Statistically, their near-symmetrical shape minimizes boundary effects and enhances signal reconstruction accuracy. Symlets are therefore best applied to denoising and empirical time series analysis, where shape and local structure need to be preserved.Fig.~\ref{fig:wavelets}

\item Coiflet Wavelet- It is developed by Ronald Coifman and Ingrid Daubechies \cite{bib39}, and are orthogonal wavelets with vanishing moments for the wavelet function and scaling function. The double vanishing moment property enhances trend estimation as well as localized feature extraction. Coiflets are the most statistically efficient for bias-variance trade-off in nonparametric regression, especially in wavelet shrinkage and smoothing. Their regularity and compact support are properties that make them best suited to applications where high accuracy in decomposition and reconstruction is required.Fig.~\ref{fig:wavelets}
\end{enumerate}

 \subsection{SARIMA}\label{subsec2}
 The ARIMA model, introduced by \citet{bib29}, is a fundamental statistical method for analyzing and forecasting time series data. It combines autoregressive (AR) terms, moving average (MA) terms, and differencing to achieve stationarity, represented as ARIMA\((p,d,q)\). The AR part captures dependence on past values, the MA part models dependence on past forecast errors, and the differencing operator accounts for trends or non-stationarity. While ARIMA models are effective for many applications, they may fail when the data exhibits seasonality. To address this,SARIMA model extends ARIMA by incorporating seasonal autoregressive, seasonal differencing, and seasonal moving average components, expressed as \(SARIMA(p,d,q)(P,D,Q)_s\), where \(s\) is the seasonal period \citep{bib48}. SARIMA models are especially powerful in capturing both short-term and long-term dependencies in data with periodic fluctuations, such as hydrological, climatological, and economic time series. Model identification usually involves examining autocorrelation (ACF) and partial autocorrelation (PACF) plots, applying seasonal differencing when necessary, and using criteria such as AIC or BIC to select appropriate orders before diagnostic checking to validate adequacy \cite{bib47},\cite{bib48}.
 
 \begin{equation}
\Phi_P(L^s)\,\phi_p(L)\,(1-L)^d\,(1-L^s)^D\,y_t \;=\; c \;+\; \Theta_Q(L^s)\,\theta_q(L)\,e_t
\end{equation}

where:
\begin{itemize}
  \item \(\phi_p(L) = 1 - \phi_1 L - \cdots - \phi_p L^p\): non-seasonal AR polynomial  
  \item \(\theta_q(L) = 1 - \theta_1 L - \cdots - \theta_q L^q\): non-seasonal MA polynomial  
  \item \(\Phi_P(L^s) = 1 - \Phi_1 L^s - \cdots - \Phi_P L^{Ps}\): seasonal AR polynomial  
  \item \(\Theta_Q(L^s) = 1 - \Theta_1 L^s - \cdots - \Theta_Q L^{Qs}\): seasonal MA polynomial  
  \item \((1-L)^d\): non-seasonal differencing operator  
  \item \((1-L^s)^D\): seasonal differencing operator  
  \item \(e_t\): white noise error term  
\end{itemize}

\subsection{Transformer}

The Transformer is an encoder–decoder-based deep learning model that eliminates recurrence by leveraging a self-attention mechanism to model dependencies between input sequence elements. The input to the Transformer is first passed through an embedding layer that maps each scalar or token \( X_i \) into a high-dimensional vector space using a learnable matrix \( E \):

\begin{equation}
\text{Embedding}(X_i) = X_i \cdot E
\end{equation}

\noindent
where \( X_i \) is the input token or value and \( E \) is the embedding matrix. Since the Transformer has no intrinsic sense of sequence order, a positional encoding is added to the embeddings using sinusoidal functions. The positional encoding for even and odd indices is defined as:

\begin{equation}
\text{PE}(pos, 2i) = \sin\left(\frac{pos}{10000^{2i/d}}\right)
\end{equation}

\begin{equation}
\text{PE}(pos, 2i+1) = \cos\left(\frac{pos}{10000^{2i/d}}\right)
\end{equation}

\noindent
Here, \( pos \) is the position in the sequence, \( i \) is the dimension index, and \( d \) is the embedding dimension. These are added to the embedded inputs to yield the final input to the encoder:

\begin{equation}
X' = \text{Embedding}(X) + \text{PE}(X)
\end{equation}

\noindent
The self-attention mechanism assigns importance scores between all tokens. For each input pair \( x_i \) and \( x_j \), the dot product is computed and normalized using a softmax function:

\begin{equation}
w_{ij} = \text{softmax}(x_i^T x_j) = \frac{e^{x_i^T x_j}}{\sum_j e^{x_i^T x_j}}
\end{equation}

\noindent
where \( w_{ij} \) represents the attention weight between the \( i \)-th and \( j \)-th tokens. Using these weights, the updated representation \( z_i \) is computed as a weighted sum of all values:

\begin{equation}
z_i = \sum_{j=1}^{n} w_{ij} x_j
\end{equation}

\noindent
To generalize this, the input is projected into Query (\( Q \)), Key (\( K \)), and Value (\( V \)) vectors using learned projection matrices:

\begin{equation}
q_i = W^Q x_i, \quad k_i = W^K x_i, \quad v_i = W^V x_i
\end{equation}

\noindent
Here, \( W^Q \), \( W^K \), and \( W^V \) are learnable matrices specific to each projection. The scaled dot-product attention is then computed as:

\begin{equation}
z_i = \sum_j \text{softmax}\left(\frac{q_i^T k_j}{\sqrt{d_k}}\right) v_j
\end{equation}

\noindent
where \( d_k \) is the dimension of the key vectors, used for scaling to control gradient magnitude. This can be rewritten in matrix form as:

\begin{equation}
Z = \text{softmax}\left(\frac{QK^T}{\sqrt{d_k}}\right)V
\end{equation}

\noindent
This attention mechanism is extended using multiple heads to form Multi-Head Attention, allowing the model to attend to information from different representation subspaces. After attention, the model applies a fully connected position-wise feedforward network:

\begin{equation}
\text{FFN}(x) = \max(0, xW_1 + b_1) W_2 + b_2
\end{equation}

\noindent
where \( W_1 \), \( W_2 \) are weight matrices, and \( b_1 \), \( b_2 \) are biases. This enables non-linear transformations at each position. To facilitate training, residual connections are used around each sublayer:

\begin{equation}
\text{Output} = \text{Sublayer}(x) + x
\end{equation}

\noindent
and the result is passed through layer normalization:

\begin{equation}
\text{LayerNorm}(x) = \gamma \cdot \frac{x - \mu}{\sigma + \epsilon} + \beta
\end{equation}

\noindent
where \( \mu \) and \( \sigma \) are the mean and standard deviation of the input, and \( \gamma \), \( \beta \) are learnable scale and shift parameters. In the decoder, masked self-attention is applied to prevent access to future positions. The final output of the decoder is projected through a linear transformation:

\begin{equation}
Y = \text{DecoderOutput} \cdot W + b
\end{equation}

\noindent
where \( W \) and \( b \) are learnable output weights and bias. Finally, a softmax function converts logits into a probability distribution:

\begin{equation}
\text{Softmax}(z_i) = \frac{e^{z_i}}{\sum_j e^{z_j}}
\end{equation}

 where $z_i$ is the ith element of the output vector from the linear layer, $e^{z_i}$ is the exponential function applied to  $z_i$ and $\sum_j e^{z_j}$ is the
sum of the exponentials of all elements 
in the output vector.

Transformer is a neural network architecture developed to be used in tasks of natural language processing such as machine translation. It has an encoder to process input and a decoder to produce output, both of which consist of stacked layers. Its self-attention mechanism enables the model to detect intricate long-range dependencies in sequential data. For a time series model, only the encoder part is necessary and Transformers are versatile and can handle both univariate as well as multivariate time series.For our study, we have used the transformer mechanism by and the steps are summarised below \cite{bib28}-

 The input to the model is a sequence of original time series data represented as:
 \begin{equation}
R = \{r_1, r_2, ..., r_T\},
\end{equation}
where \( T \) denotes the total number of time steps. The objective is to predict the series for the next \( h \) time steps, resulting in a forecasted sequence:
\begin{equation}
    \hat{R} = \{\hat{r}_{T+1}, \hat{r}_{T+2}, ..., \hat{r}_{T+h}\}.
\end{equation}

 The rainfall series is then divided into a training set \( R_{\text{train}} \) and a testing set \( R_{\text{test}} \).

To prepare the data for supervised learning, a sliding window approach is adopted. For each time step \( t \), the input sequence is constructed as:
\begin{equation}
  X = \{x_{t-w}, x_{t-w+1}, ..., x_{t-1}\},  
\end{equation}

and the corresponding output (target) sequence is given by:
\begin{equation}
    Y = \{x_t, x_{t+1}, ..., x_{t+k}\},
\end{equation}

where \( w \) is the input window size and \( k \) is the prediction horizon.

The Transformer model architecture is governed by several key hyperparameters, including:
\begin{itemize}
    \item Number of encoder layers: \( N \),
    \item Embedding dimensionality: \( d_{\text{model}} \),
    \item Feedforward network size: \( d_{\text{ff}} \),
    \item Number of attention heads: \( H \),
    \item Number of units in the multilayer perceptron (MLP): \( M \),
    \item Dropout rate: \( \delta \).
\end{itemize}

Each input sequence is passed through an embedding layer to project it into a higher-dimensional feature space. To preserve temporal ordering, sinusoidal positional encodings are added to the embeddings, yielding:
\begin{equation}
    PE(X) = E(X) + \text{positional\_encoding}(X)
\end{equation}

The encoded input is then passed through a stack of \( N \) Transformer encoder layers. Each encoder layer consists of a multi-head self-attention mechanism followed by a feedforward neural network, with both components incorporating residual connections and layer normalization. The feedforward network (FNN) is defined as:
\begin{equation}
    \text{FFN}(x) = \max(0, xW_1 + b_1)W_2 + b_2
\end{equation}

The final output of the encoder stack captures the learned representation of the input sequence and is used to generate forecasts for the desired future time steps.

\subsection{Model Validation Metrics}\label{subsec:metrics}

Validation of predictive accuracy is crucial for selecting the most suitable forecasting model. In this study, multiple statistical performance metrics are employed to ensure robustness and comparability.

\begin{itemize}
    \item \textbf{Root Mean Square Error (RMSE):}  
    RMSE measures the standard deviation of the residuals (prediction errors) and indicates how closely predictions approximate the actual values:
    \begin{equation}
        RMSE = \sqrt{\frac{1}{n} \sum_{i=1}^{n} (y_i - \hat{y}_i)^2}.
    \end{equation}
    A lower RMSE reflects higher predictive accuracy.
    
    \item \textbf{Mean Absolute Error (MAE):}  
    MAE computes the average absolute difference between observed and predicted values, treating over- and under-predictions equally:
    \begin{equation}
        MAE = \frac{1}{n} \sum_{i=1}^{n} |y_i - \hat{y}_i|.
    \end{equation}
    This metric provides an intuitive measure of average prediction error in the original units.
    
    \item \textbf{Symmetric Mean Absolute Percentage Error (SMAPE):}  
    SMAPE expresses errors as percentages of the average of observed and predicted values, making it scale-independent:
    \begin{equation}
        SMAPE = \frac{100}{n} \sum_{i=1}^{n} \frac{|y_i - \hat{y}_i|}{(|y_i| + |\hat{y}_i|)/2}.
    \end{equation}
    Its symmetry ensures equal penalization of over- and under-predictions.
    
    \item \textbf{Willmott’s Index of Agreement ($d$):}  
    This index measures the degree of agreement between predictions and observations, ranging from 0 (no agreement) to 1 (perfect match). 
    \begin{equation}
        d = 1 - \frac{\sum_{i=1}^{n}(y_i - \hat{y}_i)^2}{\sum_{i=1}^{n}\left(|\hat{y}_i - \bar{y}| + |y_i - \bar{y}|\right)^2}.
    \end{equation}
    
    \item \textbf{Skill Score (SS):}  
    SS compares the predictive skill of a model relative to a reference (often climatology or persistence). Positive values close to 1 indicate superior performance, while negative values indicate poorer skill.
    
    \item \textbf{Percent Bias (PBIAS):}  
    PBIAS quantifies systematic overestimation or underestimation:
    \begin{equation}
        PBIAS = 100 \times \frac{\sum_{i=1}^{n}(y_i - \hat{y}_i)}{\sum_{i=1}^{n} y_i}.
    \end{equation}
    Positive values imply underestimation, whereas negative values imply overestimation.
    
    \item \textbf{Explained Variance (EV):}  
    EV evaluates the proportion of variance in the observed data explained by the model:
    \begin{equation}
        EV = 1 - \frac{\mathrm{Var}(y_i - \hat{y}_i)}{\mathrm{Var}(y_i)}.
    \end{equation}
    Values close to 1 denote strong explanatory power.
    
    \item \textbf{Legates \& McCabe’s Efficiency Index ($E_1$):}  
    Unlike RMSE, $E_1$ is based on absolute errors and penalizes extreme deviations less severely:
    \begin{equation}
        E_1 = 1 - \frac{\sum_{i=1}^{n} |y_i - \hat{y}_i|}{\sum_{i=1}^{n} |y_i - \bar{y}|}.
    \end{equation}
    A value of 1 represents perfect agreement, while values $\leq 0$ indicate performance no better than the mean.
\end{itemize}

Together, these metrics provide a comprehensive evaluation, capturing error magnitude (RMSE, MAE), relative accuracy (SMAPE, PBIAS), agreement (d, $E_1$), explanatory power (EV), and comparative skill (SS).

\subsection{Hybrid Wavelet--SARIMA--Transformer Algorithm}\label{subsec2d}

The novelty of the proposed WST framework lies in \textbf{synergistically integrating MODWT with SARIMA and Transformer}. MODWT separates the time series into scale-specific components, SARIMA captures linear seasonality, and the Transformer models nonlinear dynamics. This dual modeling enhances robustness against noise and improves generalization compared to standalone approaches. The steps of the proposed hybrid algorithm are summarized as follows:

\begin{enumerate}
  \item \textbf{Input:} Time series $X_t$, forecast horizon $h$.
  \item Apply MODWT to decompose $X_t$ into smooth ($V_t$) and detail ($W_t$) components.
  \item Perform Tsay test to classify components into linear and nonlinear.
  \item Fit SARIMA to linear; fit Transformer to nonlinear.
  \item Reconstruct predicted time series using IMODWT.
  \item \textbf{Output:} Predicted Time series $\hat{X}_t$ with Forecasting .
  \item Evaluate performance using RMSE, MAE, SMAPE, Willmott’s $d$, Skill Score, PBIAS, EV, $E_1$, correlation $r$, Std.~Dev, and centered RMSE.
\end{enumerate}

\section{Data Used}\label{sec:data}

The Indian subcontinent is climatologically classified by the India Meteorological Department (IMD) into 36 meteorological subdivisions, of which 34 are land-based and 2 correspond to island regions. Each subdivision represents a zone with broadly homogeneous climatic and rainfall characteristics (\url{https://mausam.imd.gov.in/}). 

For this study, we focus on the Northeastern Region (NER) of India \cite{bib52}, which is highly sensitive to monsoon variability and extreme rainfall events. The NER consists of the following five meteorological subdivisions:

\begin{enumerate}
    \item Assam and Meghalaya (ASML),
    \item Arunachal Pradesh (ARP),
    \item Nagaland, Manipur, Mizoram, and Tripura (NMMT),
    \item Gangetic West Bengal (GWB),
    \item Sub-Himalayan West Bengal and Sikkim (SHWBS).
\end{enumerate}

Monthly rainfall data for these subdivisions were obtained from the India Meteorological Department (IMD), Pune (\url{https://imdpune.gov.in/}). The dataset spans a period of 53 years, from January 1971 to December 2023, providing a sufficiently long record for robust statistical analysis of seasonal and interannual variability.

Among the 36 IMD-defined subdivisions, these five were selected as they represent the northeastern sector of India, a region characterized by complex topography, high rainfall variability, and pronounced sensitivity to monsoon fluctuations. This makes it an ideal case study for validating the proposed hybrid modeling framework.
\begin{figure}[H]
    \centering
    \includegraphics[scale=0.5]{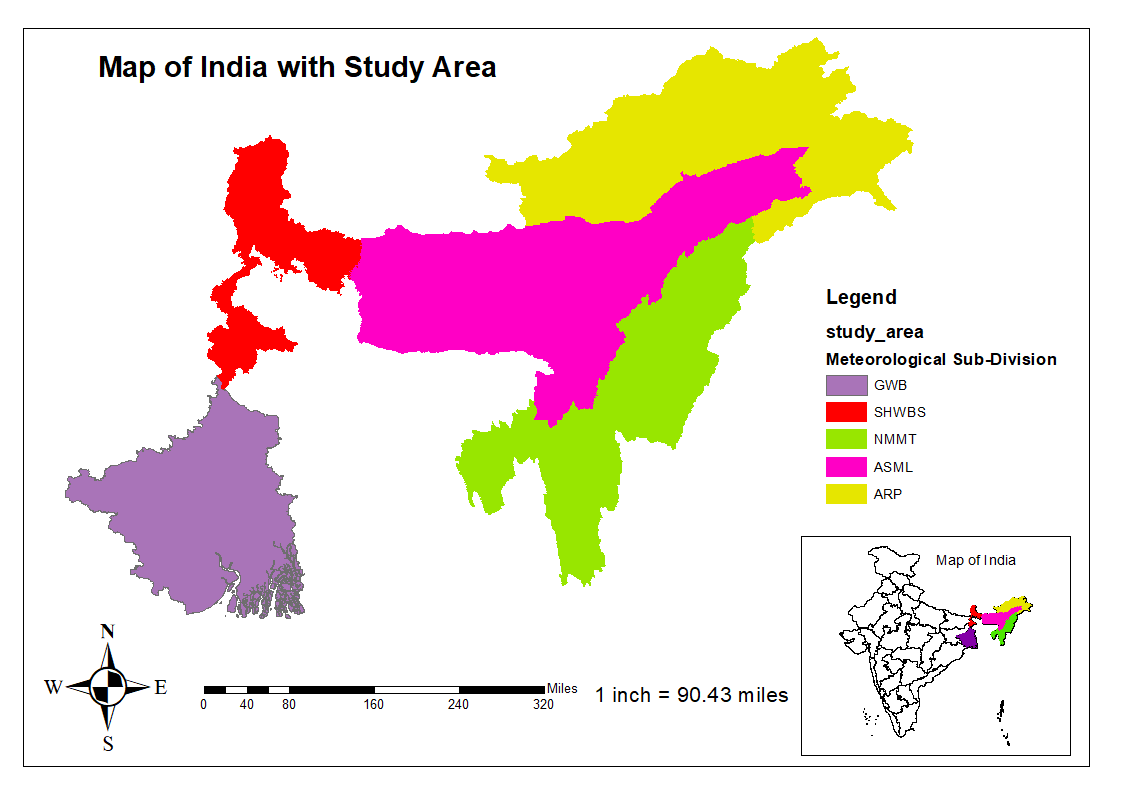}
    \caption{ Map of the northeast region of India showing different meteorological subdivisions.}
   
\end{figure}

\section{Results and Discussion}
This section presents the modeling results for monthly rainfall in the Northeastern Region (NER) of India, which comprises five meteorological subdivisions. The analysis follows a systematic approach, beginning with descriptive statistics, time series visualization, wavelet-based decomposition, model evaluation, and concluding with forecasting performance. The proposed Wavelet–SARIMA–Transformer (W-ST) hybrid framework is benchmarked against alternative models to establish its predictive superiority.
\subsection{Descriptive Statistics of Rainfall Data}
This subsection presents the descriptive statistics of the monthly rainfall data (1971–2023) for the five selected meteorological subdivisions of Northeast India. A total of 636 monthly observations were considered for each subdivision, with 509 (80\%) used for model training and 127 (20\%) reserved for testing. Tab.~\ref{tab:desc_stats} 

The statistical characteristics—including mean, standard deviation, minimum, maximum, median, skewness, and kurtosis—are summarized for both training and testing subsets. These metrics provide insight into the underlying distributional properties, highlighting the high variability and skewness typical of hydrometeorological series. 
The descriptive statistics in Table~\ref{tab:desc_stats} reveal high spatial heterogeneity. ARP and SHWBS exhibit the largest mean monthly rainfall (227.93\,mm and 222.02\,mm, respectively) with substantial variability (SDs of 188.85\,mm and 226.33\,mm). ASML witnessed high-rainfall (mean 207.99\,mm; SD 196.97\,mm) and shows the largest extreme (Max $\approx$ 1008.8\,mm), pointing to occasional intense monsoon bursts. In contrast, GWB has the lowest mean (131.10\,mm) yet retains high dispersion (SD 140.17\,mm), indicating sporadic extremes. Positive skewness across regions (about 0.6–1.0) indicates right-tailed distributions dominated by episodic heavy rainfall; kurtosis near zero (slightly negative in places) suggests near-mesokurtic behavior. Together, these features justify a hybrid approach: linear seasonal structure coexists with nonlinear, intermittent variability.

\begin{table}[h]
\caption{Descriptive statistics of monthly rainfall (1971–2023) across five meteorological subdivisions of Northeast India.}
\label{tab:desc_stats}
\begin{tabular}{@{}p{1.5cm}llllllllll@{}}
\toprule
Subdivisions & Data & Count & Mean & Std Dev & Min & Median & Max & Skewness & Kurtosis \\
\midrule
ARP & Total    & 636 & 227.93 & 188.85 & 0.3 & 194.8 & 834.8 & 0.80 & -0.07 \\
    & Training & 509 & 233.97 & 192.46 & 0.3 & 200.1 & 834.8 & 0.81 & -0.07 \\
    & Testing  & 127 & 203.71 & 172.24 & 0.8 & 163.7 & 683.5 & 0.68 & -0.48 \\
\midrule
ASML & Total    & 636 & 207.99 & 196.97 & 0.2 & 163.55 & 1008.8 & 0.83 & 0.01 \\
     & Training & 509 & 209.79 & 197.48 & 0.2 & 163.6  & 1008.8 & 0.82 & -0.01 \\
     & Testing  & 127 & 200.79 & 195.54 & 0.3 & 145.9  & 858.1  & 0.86 & 0.09 \\
\midrule
NMMT & Total    & 636 & 180.11 & 163.38 & 0   & 151.35 & 978.7 & 0.71 & 0.01 \\
     & Training & 509 & 187.58 & 168.84 & 0   & 155.1  & 978.7 & 0.68 & -0.05 \\
     & Testing  & 127 & 150.17 & 135.89 & 0.1 & 113.3  & 569.5 & 0.60 & -0.63 \\
\midrule
GWB  & Total    & 636 & 131.10 & 140.17 & 0   & 69.4  & 609.8 & 1.04 & 0.23 \\
     & Training & 509 & 132.08 & 141.94 & 0   & 67.8  & 604.4 & 1.04 & 0.16 \\
     & Testing  & 127 & 127.18 & 133.29 & 0   & 76.5  & 609.8 & 1.04 & 0.56 \\
\midrule
SHWBS & Total    & 636 & 222.02 & 226.33 & 0   & 131.1 & 911.5 & 0.89 & -0.26 \\
      & Training & 509 & 222.53 & 228.07 & 0.1 & 128.7 & 911.5 & 0.92 & -0.20 \\
      & Testing  & 127 & 219.94 & 220.09 & 0   & 146.7 & 801.3 & 0.78 & -0.54 \\
\botrule
\end{tabular}
\end{table}

\begin{figure}[H]
    \centering
    \includegraphics[scale=0.3]{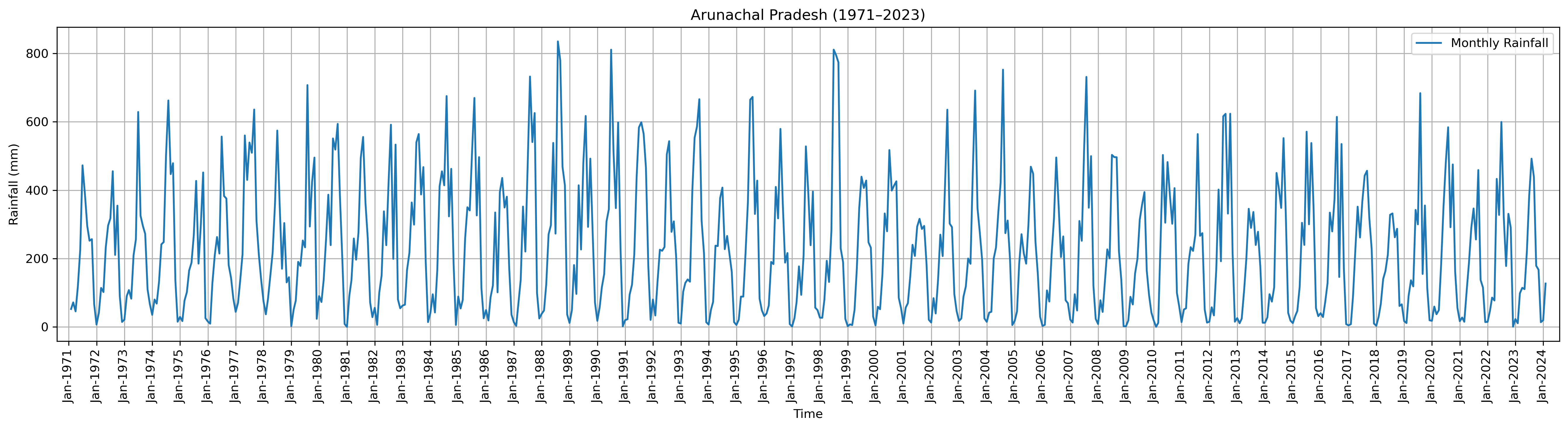}
    \caption{Time series plot of monthly rainfall (1971–2023) for Arunachal Pradesh (ARP).}
    \label{fig:rainfall_ap}
\end{figure}

\begin{figure}[H]
    \centering
    \includegraphics[scale=0.3]{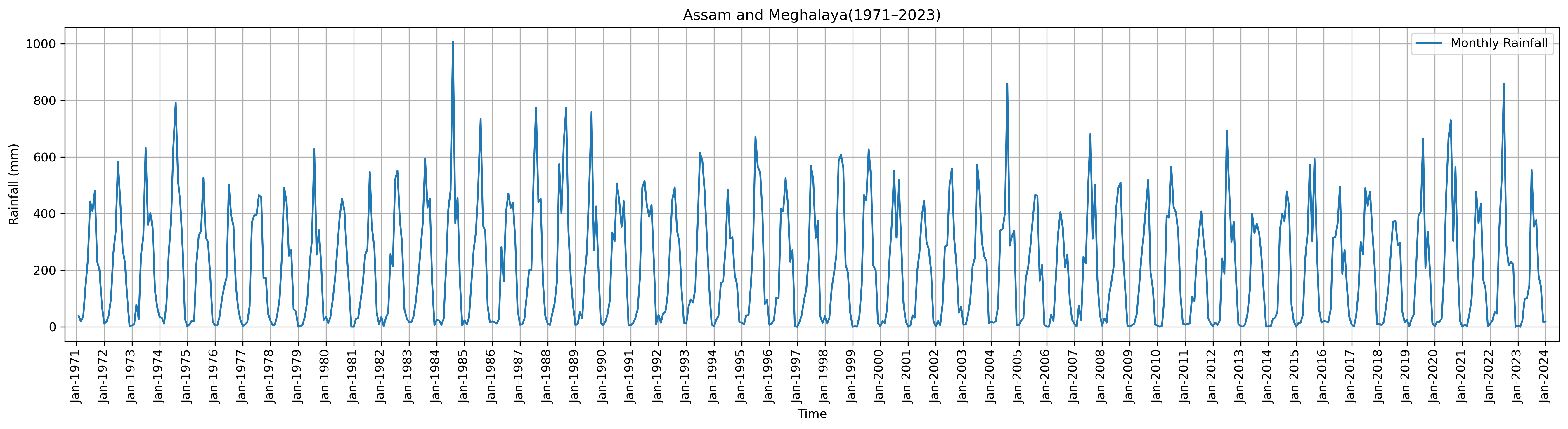}
    \caption{Time series plot of monthly rainfall (1971–2023) for Assam and Meghalaya (ASML).}
    \label{fig:rainfall_asml}
\end{figure}

\begin{figure}[H]
    \centering
    \includegraphics[scale=0.3]{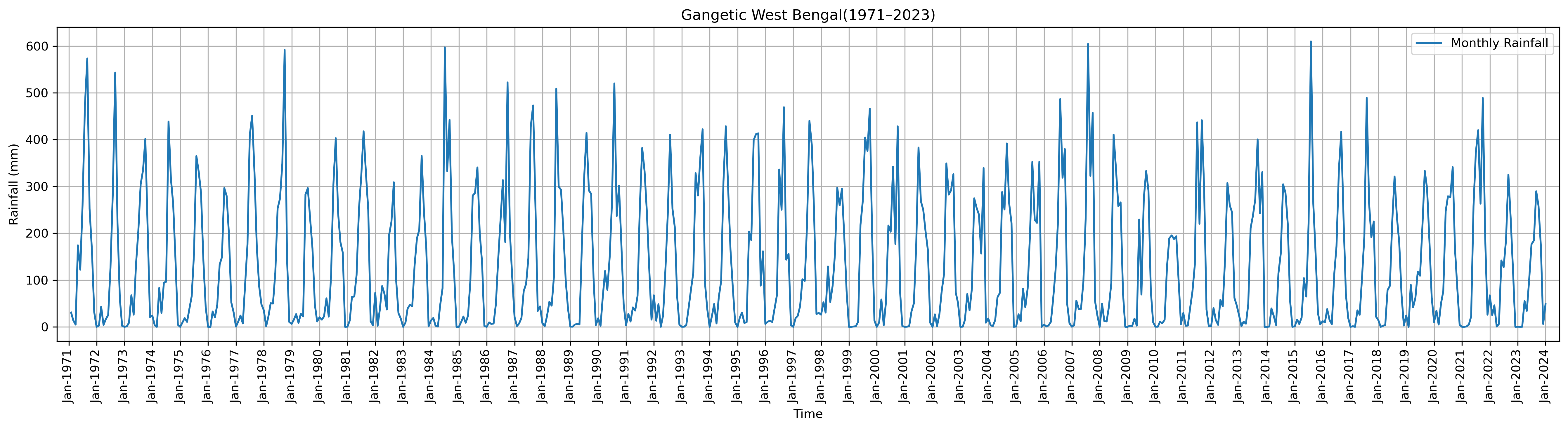}
    \caption{Time series plot of monthly rainfall (1971–2023) for Gangetic West Bengal (GWB).}
    \label{fig:rainfall_gwb}
\end{figure}

\begin{figure}[H]
    \centering
    \includegraphics[scale=0.3]{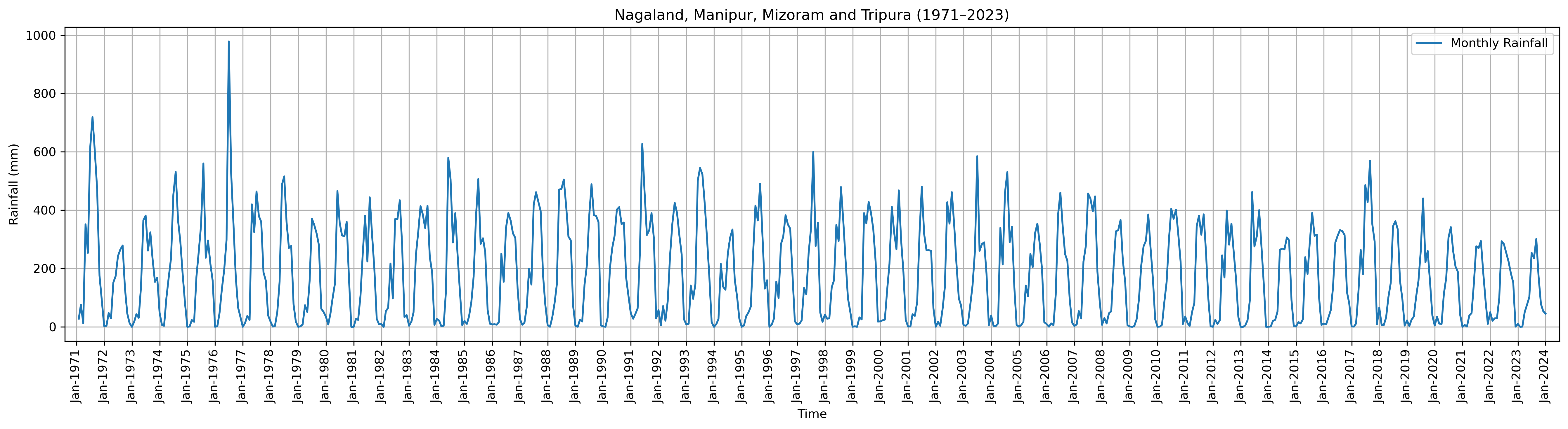}
    \caption{Time series plot of monthly rainfall (1971–2023) for Nagaland–Manipur–Mizoram–Tripura (NMMT).}
    \label{fig:rainfall_nmmt}
\end{figure}

\begin{figure}[H]
    \centering
    \includegraphics[scale=0.3]{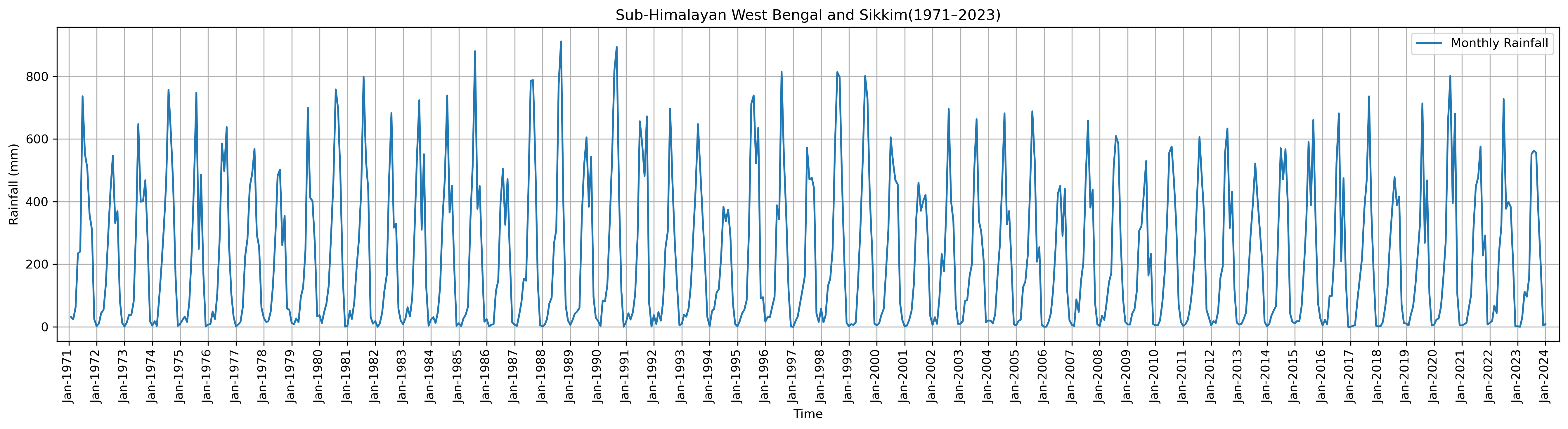}
    \caption{Time series plot of monthly rainfall (1971–2023) for Sub-Himalayan West Bengal and Sikkim (SHWBS).}
    \label{fig:rainfall_shwbs}
\end{figure}

The time series plots (Fig.~\ref{fig:rainfall_ap}–\ref{fig:rainfall_shwbs}) reveal complex rainfall dynamics characterized by strong seasonal fluctuations, abrupt breaks, and anomalies. ASML and SHWBS consistently exhibit higher rainfall magnitudes, whereas GWB and NMMT show more irregular and volatile rainfall patterns.

\subsection{Wavelet Decomposition and Model Selection}
For our dataset of length $N = 636$, the maximum permissible level of wavelet decomposition is determined by  
\begin{equation}
J_{\max} = \left\lfloor \log_{2}\!\left(\frac{N}{L - 1}\right) \right\rfloor ,
\end{equation}
where $L$ denotes the length of the wavelet filter. Applying this criterion, the Haar wavelet ($L = 2$) permits up to $J_{\max} = 9$ decomposition levels, while both Daubechies-4 (db4, $L = 8$) and Symlet-4 (sym4, $L = 8$) allow a maximum of $J_{\max} = 7$ levels, and Coiflet-3 (coif3, $L = 18$) supports up to $J_{\max} = 5$ levels.  

Although these values represent the theoretical upper bound, in practice decompositions were carried out between levels 5 and 9. Consistent with prior findings that emphasize the trade-off between detail preservation and noise amplification, the most reliable predictive performance was obtained at level 8, which was therefore selected for further analysis. This choice aligns with the approach of \cite{bib53}, who highlighted the importance of jointly selecting an appropriate mother wavelet and optimum decomposition depth to enhance forecasting accuracy.
Following MODWT decomposition, we evaluated four wavelet families—Haar, Daubechies (Db4), Symlet (Sym4), and Coiflet (Coif4)—with the selection for each region guided by the balance between detail preservation and smoothness. In practice, Db4 performed best in regions with gradual rainfall transitions, whereas Haar was advantageous where abrupt shifts were frequent, consistent with the hydrometeorological behavior of the series.
To route components to the appropriate learners, we applied the Tsay nonlinearity test to each decomposed subseries. Components exhibiting seasonality and linear dependence were modeled via SARIMA; components with nonlinear dependence were modeled via a Transformer encoder. This design operationalizes the complementary strengths of stochastic linear models (seasonality, short-to-medium memory) and attention-based deep learners (long-range and nonlinear dependencies), while wavelets provide multiresolution alignment of signals and reduce aliasing via shift-invariant decomposition.

\subsection{Hyperparameter Tuning of the Transformer Architecture}
A crucial stage in the development of the proposed hybrid framework was the optimization of the Transformer component. The search strategy focused on identifying an architecture that balances learning capacity with generalization, while avoiding overfitting given the moderate sample size of rainfall series. After systematic experimentation, the best-performing configuration was determined as follows: two Transformer blocks, head size = 128, feed-forward dimension ($d_{ff}$) = 4, four attention heads, $mlp\_units = 64$, dropout rate = 0.25, and $mlp\_dropout = 0.40$.  

Training dynamics were also optimized by exploring batch sizes (16, 32, 64), epoch counts (50, 100, 200), and input sequence lengths (8, 12, 16 months). The most effective setup was achieved with a batch size of 32, an input sequence length of 12 months, and 100 training epochs with early stopping to prevent overfitting. This configuration offered a stable convergence profile, minimized validation error, and achieved the most favorable balance between predictive accuracy and computational efficiency.  

These results emphasize that model accuracy in nonlinear sequence forecasting depends not only on network depth or complexity but also on carefully chosen hyperparameters that reflect the temporal resolution of the data. The identified configuration was therefore employed in the W-ST model across all subdivisions.

\subsection{Comparative Accuracy Across Models}
We compared nine alternatives: stand-alone SARIMA and Transformer; two-stage hybrids (Wavelet–SARIMA, Wavelet–Transformer) under four wavelets; and the proposed three-stage hybrid (Wavelet–SARIMA–Transformer, W-ST) under four wavelets. Accuracy was assessed using RMSE, MAE, SMAPE, Willmott’s index $d$, Skill Score (SS), Percent Bias (PBIAS), Explained Variance (EV), and Legates–McCabe $E_1$. Tab.~\ref{tab:accuracy} 

\begin{sidewaystable}
\caption{Accuracy measures of different models across sub-regions (AP, ASMG, GWB, NAGA, SHWB).}
\label{tab:accuracy}
\begin{tabular*}{\textheight}{@{\extracolsep\fill}p{.7cm}p{1.8cm}p{1.2cm}p{1.3cm}cccccccccccc}
\toprule%
\textbf{Sub} & \textbf{Accuracy measures} & \textbf{SARIMA} & \textbf{Transformer} &
\multicolumn{4}{c}{\textbf{Wavelet-SARIMA}} &
\multicolumn{4}{c}{\textbf{Wavelet-Transformer}} &
\multicolumn{4}{c}{\textbf{Wavelet-SARIMA-Transformer}} \\
\cmidrule(lr){4-4} \cmidrule(lr){5-8} \cmidrule(lr){9-12} \cmidrule(lr){13-16}
& & & & Haar & Daube & sym & coif & Haar & Daube & sym & coif & Haar & Daube & sym & coif \\
\midrule
ARP & RMSE & 98.731 & 112.82 & 99.653 & 134.00 & 138.66 & 196.07 & 88.058 & 130.80 & 135.31 & 189.36 & 82.598 & 129.70 & 133.69 & 191.67 \\
   & MAE  & 75.048 & 80.700 & 74.049 & 100.28 & 97.791 & 148.38 & 67.296 & 95.782 & 92.253 & 141.83 & 62.226 & 96.074 & 93.793 & 146.94 \\
   & SMAPE(\%) & 53.24 & 57.97  & 49.49 & 63.19 & 61.52 & 80.31 & 51.33 & 66.42 & 65.65 & 80.37 & 47.15 & 63.40 & 65.73 & 81.77 \\
   & Wilmott’s Index & 0.8761 & 0.8580 & 0.8834 & 0.7911 & 0.778 & 0.5922 & 0.9134 & 0.8134 & 0.7975 & 0.5955 & 0.9254 & 0.8164 & 0.8027 & 0.5922 \\
   & Skill Score & 0.6680 & 0.5855 & 0.6617 & 0.3883 & 0.3451 & -0.3095 & 0.7393 & 0.4247 & 0.3843 & -0.2057 & 0.7706 & 0.4343 & 0.399 & -0.2353 \\
   & Abs. Bias(\%) & 2.416 & 0.4427 & 0.537 & 10.186 & 1.25 & 19.10 & 1.884 & 6.91 & 5.53 & 11.657 & 0.649 & 10.31 & 1.36 & 18.49 \\
   & Expl. Var. & 0.6688 & 0.5855  & 0.6618 & 0.4032 & 0.3453 & -0.2574 & 0.7398 & 0.4316 & 0.3887 & -0.1862 & 0.7707 & 0.4496 & 0.3992 & -0.1862 \\
   & Legates \& McCabe & 0.7021 & 0.7042 & 0.7178 & 0.6241 & 0.6375 & 0.4773 & 0.7495 & 0.6524 & 0.6641 & 0.4802 & 0.7703 & 0.6509 & 0.6576 & 0.4658 \\
\midrule
ASML & RMSE & 95.136 & 114.12 & 152.71 & 138.81 & 148.85 & 212.20 & 92.552 & 141.80 & 151.65 & 213.70 & 85.256 & 143.47 & 152.20 & 215.42 \\
     & MAE  & 64.616 & 81.848 & 118.71 & 97.034 & 101.12 & 152.39 & 62.480 & 98.679 & 101.18 & 152.95 & 55.963 & 99.411 & 99.268 & 153.97 \\
     & SMAPE(\%) & 54.03 & 71.18 & 83.66 & 68.50 & 81.15 & 97.14 & 65.18 & 79.63 & 77.52 & 93.10 & 59.33 & 76.77 & 79.21 & 94.63 \\
     & Wilmott’s Index & 0.921 & 0.8930 & 0.6744 & 0.8254 & 0.7971 & 0.6005 & 0.9294 & 0.8245 & 0.7951 & 0.5676 & 0.9419 & 0.8208 & 0.7955 & 0.5642 \\
     & Skill Score & 0.7615 & 0.6676 & 0.3854 & 0.4921 & 0.4161 & -0.1868 & 0.7796 & 0.4825 & 0.4081 & -0.1753 & 0.8129 & 0.4702 & 0.4038 & -0.1942 \\
     & Abs. Bias(\%) & 2.86 & 0.1139 & 3.39 & 4.23 & 2.6213 & 7.2083 & 2.2533 & 3.5989 & 3.0092 & 5.6916 & 3.8682 & 3.90 & 3.42 & 5.60 \\
     & Expl. Var. & 0.7623 & 0.6676 & 0.3866 & 0.494 & 0.4168 & -0.1812 & 0.7801 & 0.4839 & 0.4091 & -0.1717 & 0.8146 & 0.4719 & 0.4051 & -0.1908 \\
     & Legates \& McCabe & 0.7863 & 0.7399 & 0.4871 & 0.6748 & 0.6636 & 0.5102 & 0.7998 & 0.6809 & 0.6702 & 0.4925 & 0.8614 & 0.6788 & 0.6786 & 0.4924 \\
\midrule
NMMT & RMSE & 65.159 & 62.422 & 96.291 & 116.82 & 132.33 & 155.78 & 62.526 & 104.35 & 106.38 & 157.23 & 59.074 & 105.04 & 106.65 & 158.39 \\
     & MAE  & 50.165 & 42.769 & 80.993 & 87.442 & 117.14 & 119.38 & 49.563 & 78.889 & 76.688 & 122.49 & 46.906 & 79.799 & 77.201 & 122.84 \\
     & SMAPE(\%) & 58.28 & 56.55 & 78.40 & 79.66 & 92.36 & 93.66 & 62.55 & 79.92 & 78.62 & 98.09 & 62.21 & 78.85 & 80.02 & 95.02 \\
     & Wilmott’s Index & 0.9387 & 0.9350 & 0.7671 & 0.7677 & 0.2778 & 0.6607 & 0.9362 & 0.8348 & 0.8238 & 0.6361 & 0.9427 & 0.8326 & 0.8216 & 0.6308 \\
     & Skill Score & 0.7758 & 0.7844 & 0.5103 & 0.2792 & 0.0751 & -0.2817 & 0.8020 & 0.4483 & 0.4267 & -0.2524 & 0.8232 & 0.4410 & 0.4237 & -0.2709 \\
     & Abs. Bias(\%) & 15.89 & 1.8896 & 13.61 & 21.20 & 6.0000 & 29.034 & 6.419 & 18.391 & 5.2154 & 23.383 & 5.171 & 18.935 & 4.4127 & 25.893 \\
     & Expl. Var. & 0.8068 & 0.7849 & 0.5331 & 0.3345 & 0.0795 & -0.1780 & 0.8071 & 0.4904 & 0.4301 & -0.1845 & 0.8265 & 0.4855 & 0.4262 & -0.1876 \\
     & Legates \& McCabe & 0.7901 & 0.8060 & 0.5415 & 0.6031 & 0.1558 & 0.5141 & 0.7767 & 0.6585 & 0.6656 & 0.4884 & 0.7882 & 0.6546 & 0.6634 & 0.4878 \\
\midrule
GWB & RMSE & 66.080 & 75.364 & 91.261 & 129.27 & 126.02 & 149.05 & 66.344 & 102.01 & 107.75 & 137.69 & 61.924 & 101.79 & 107.49 & 137.32 \\
    & MAE  & 44.022 & 50.651 & 68.671 & 101.50 & 98.614 & 118.15 & 45.480 & 68.017 & 70.423 & 99.960 & 40.346 & 67.975 & 69.611 & 99.194 \\
    & SMAPE(\%) & 69.10 & 78.30 & 83.93 & 100.13 & 100.35 & 111.26 & 69.01 & 84.12 & 87.04 & 99.73 & 67.87 & 82.73 & 85.92 & 96.88 \\
    & Wilmott’s Index & 0.9236 & 0.8909 & 0.7683 & 0.5689 & 0.5533 & 0.4564 & 0.9189 & 0.8101 & 0.7863 & 0.6442 & 0.9306 & 0.8105 & 0.7873 & 0.6509 \\
    & Skill Score & 0.7511 & 0.6695 & 0.5253 & 0.0473 & 0.0947 & -0.2664 & 0.7598 & 0.4322 & 0.3664 & -0.0345 & 0.7908 & 0.4346 & 0.3695 & -0.0290 \\
    & Abs. Bias(\%) & 1.03 & 4.9547 & 1.07 & 3.72 & 0.5298 & 6.5400 & 0.4998 & 4.0715 & 1.1283 & 5.6320 & 4.7761 & 3.1777 & 0.3594 & 5.6817 \\
    & Expl. Var. & 0.7512 & 0.6718 & 0.5254 & 0.0486 & 0.0948 & -0.2624 & 0.7599 & 0.4337 & 0.3666 & -0.0316 & 0.7929 & 0.4356 & 0.3695 & -0.0260 \\
    & Legates \& McCabe & 0.7957 & 0.7503 & 0.5859 & 0.4267 & 0.4166 & 0.3551 & 0.7782 & 0.6724 & 0.6597 & 0.5158 & 0.8406 & 0.6724 & 0.6631 & 0.5222 \\
\midrule

SHWBS & RMSE & 86.464 & 114.31 & 111.02 & 155.61 & 165.27 & 230.72 & 90.186 & 149.36 & 158.79 & 232.16 & 84.716 & 153.18 & 158.78 & 232.13 \\
     & MAE  & 57.292 & 75.741 & 77.505 & 106.02 & 107.72 & 160.97 & 58.086 & 96.357 & 99.177 & 165.52 & 53.054 & 99.354 & 99.401 & 165.40 \\
     & SMAPE(\%) & 47.23 & 67.57 & 63.09 & 74.90 & 72.20 & 93.22 & 57.78 & 73.79 & 75.50 & 99.63 & 54.69 & 75.47 & 75.79 & 99.69 \\
     & Wilmott’s Index & 0.9545 & 0.9157 & 0.9081 & 0.8271 & 0.8079 & 0.6435 & 0.9481 & 0.8503 & 0.8314 & 0.6161 & 0.9542 & 0.8414 & 0.9487 & 0.6162 \\
     & Skill Score & 0.8433 & 0.7350 & 0.7417 & 0.4926 & 0.4276 & -0.1155 & 0.8298 & 0.5332 & 0.4724 & -0.1278 & 0.8498 & 0.5090 & 0.8217 & -0.1274 \\
     & Abs. Bias(\%) & 1.761 & 5.0099 & 4.14 & 0.22 & 3.177 & 3.578 & 5.524 & 1.478 & 3.280 & 3.807 & 5.63 & 0.544 & 1.702 & 3.630 \\
     & Expl. Var. & 0.8437 & 0.7375 & 0.7434 & 0.4926 & 0.4286 & -0.1142 & 0.8329 & 0.5334 & 0.4735 & -0.1263 & 0.8531 & 0.5091 & 0.8220 & -0.1261 \\
     & Legates \& McCabe & 0.8421 & 0.7876 & 0.7628 & 0.6849 & 0.6833 & 0.5466 & 0.8367 & 0.7241 & 0.7166 & 0.5164 & 0.8764 & 0.7756 & 0.8680 & 0.5168 \\
\botrule
\end{tabular*}
\end{sidewaystable}

\paragraph{Error Magnitude (RMSE/MAE).}
The Haar-based W-ST [W(H)-ST] consistently yields the lowest errors. Illustratively:
(i) ARP: RMSE drops from 98.73 (SARIMA) to 82.60 [W(H)-ST], a $\sim$16\% reduction; MAE drops from 75.05 to 62.23 ($\sim$13\%). 
(ii) ASML: RMSE 95.14 $\rightarrow$ 82.883 ($\sim$12\% reduction); MAE 64.62 $\rightarrow$ 54.88 ($\sim$9\%). 
(iii) GWB: RMSE 66.08 $\rightarrow$ 61.92 ($\sim$4\%); 
(iv) NMMT: 65.16 $\rightarrow$ 59.07 ($\sim$5\%); 
(v) SHWBS/SHWB: 86.46 $\rightarrow$ 84.72 ($\sim$2\%). 
Gains are largest in higher-variance regions (ARP, ASML), and more modest where seasonality is dominant and variance is already well captured by SARIMA (SHWBS).

\paragraph{Relative and Symmetric Errors (SMAPE).}
W(H)-ST reduces SMAPE relative to SARIMA in every region (e.g., ARP: 53.24\% $\rightarrow$ 47.15\%; GWB: 69.10\% $\rightarrow$ 67.87\%), showing improved scale-free accuracy with balanced penalties for over- and under-forecasting.

\paragraph{Agreement and Skill (Willmott’s $d$, SS).}
Willmott’s $d$ improves consistently (e.g., ARP: 0.876 $\rightarrow$ 0.925; SHWB: 0.955 $\rightarrow$ 0.954–0.955 range with hybrids), indicating stronger pattern agreement with observations. Skill Score remains positive and higher for W(H)-ST across regions (e.g., ARP: SS $\approx$ 0.77 for W(H)-ST vs.\ 0.67 for SARIMA), confirming consistent improvement over the reference.

\paragraph{Bias, Variance Explained, and Robustness (PBIAS, EV, $E_1$).}
Bias magnitudes remain small under W(H)-ST (e.g., ARP PBIAS $\approx$ 0.65\%, balanced forecasts). Explained Variance (EV) and Legates–McCabe $E_1$ both increase for W(H)-ST (e.g., ARP: EV 0.669 $\rightarrow$ 0.771; $E_1$ 0.702 $\rightarrow$ 0.770), indicating better signal capture and reliability under absolute-error loss.

\subsection{Model Adequacy: Residual Diagnostics}
The adequacy of the fitted time series models was assessed using the Ljung–Box test, which evaluates whether the residuals are independently distributed. The test results presented in Table~\ref{tab:ljung} show that for Arunachal Pradesh (statistic = 17.6853, p = 0.1256), Assam and Meghalaya (20.3463, p = 0.0608), Nagaland, Manipur, Mizoram, and Tripura (20.9074, p = 0.0518), Gangetic West Bengal (20.0809, p = 0.0656), and Sub-Himalayan West Bengal with Sikkim (7.4452, p = 0.8268), the null hypothesis of no autocorrelation cannot be rejected at the 5\% significance level. These findings indicate that the residuals behave like white noise, suggesting that the models have effectively captured the underlying serial correlation structure in the data. Therefore, the fitted models can be considered statistically adequate, reliable, and suitable for forecasting and further inferential analysis.Tab.~\ref{tab:ljung}

\begin{table}[h]
\caption{Results of Ljung-Box test for residual autocorrelation across subdivisions.}
\label{tab:ljung}
\begin{tabular*}{\textwidth}{@{\extracolsep\fill}p{2cm}ccp{4cm}}
\toprule
\textbf{Subdivisions} & \textbf{Statistic} & \textbf{P-Value} & \textbf{Interpretation} \\
\midrule
AP & 17.6853 & 0.1256 & Residuals show no significant autocorrelation at 5\% level. \\
ASML & 20.3463 & 0.0608 & Residuals show no significant autocorrelation at 5\% level. \\
NMMT & 20.9074 & 0.0518 & Residuals show no significant autocorrelation at 5\% level. \\
GWB & 20.0809 & 0.0656 & Residuals show no significant autocorrelation at 5\% level. \\
SHWBS & 7.4452 & 0.8268 & Residuals show no significant autocorrelation at 5\% level. \\
\botrule
\end{tabular*}
\end{table}

\subsection{Correlation–Variance–RMSE Synthesis via Taylor Diagrams}
\begin{figure}[htbp]
\centering
\includegraphics[width=0.45\textwidth]{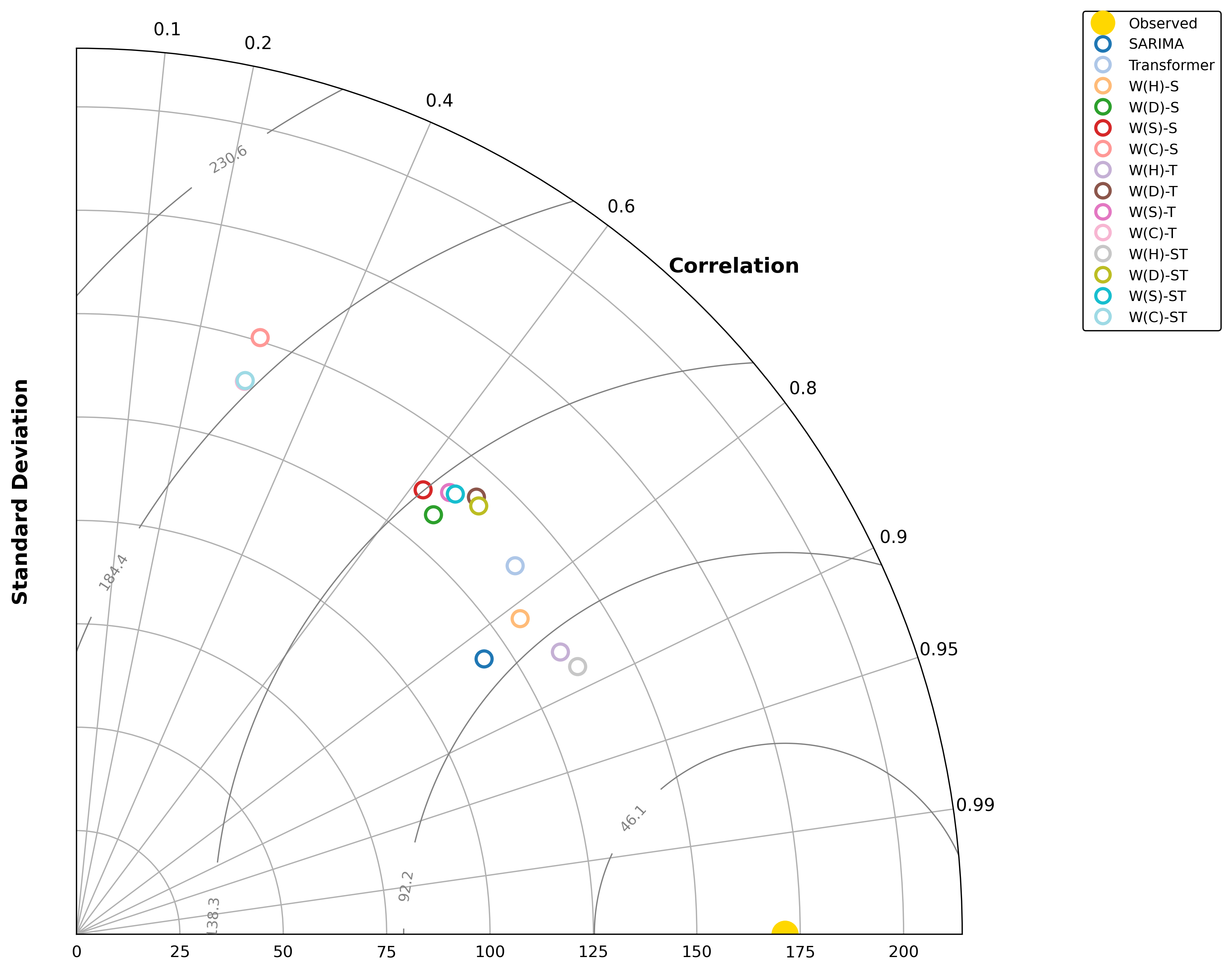}
\includegraphics[width=0.45\textwidth]{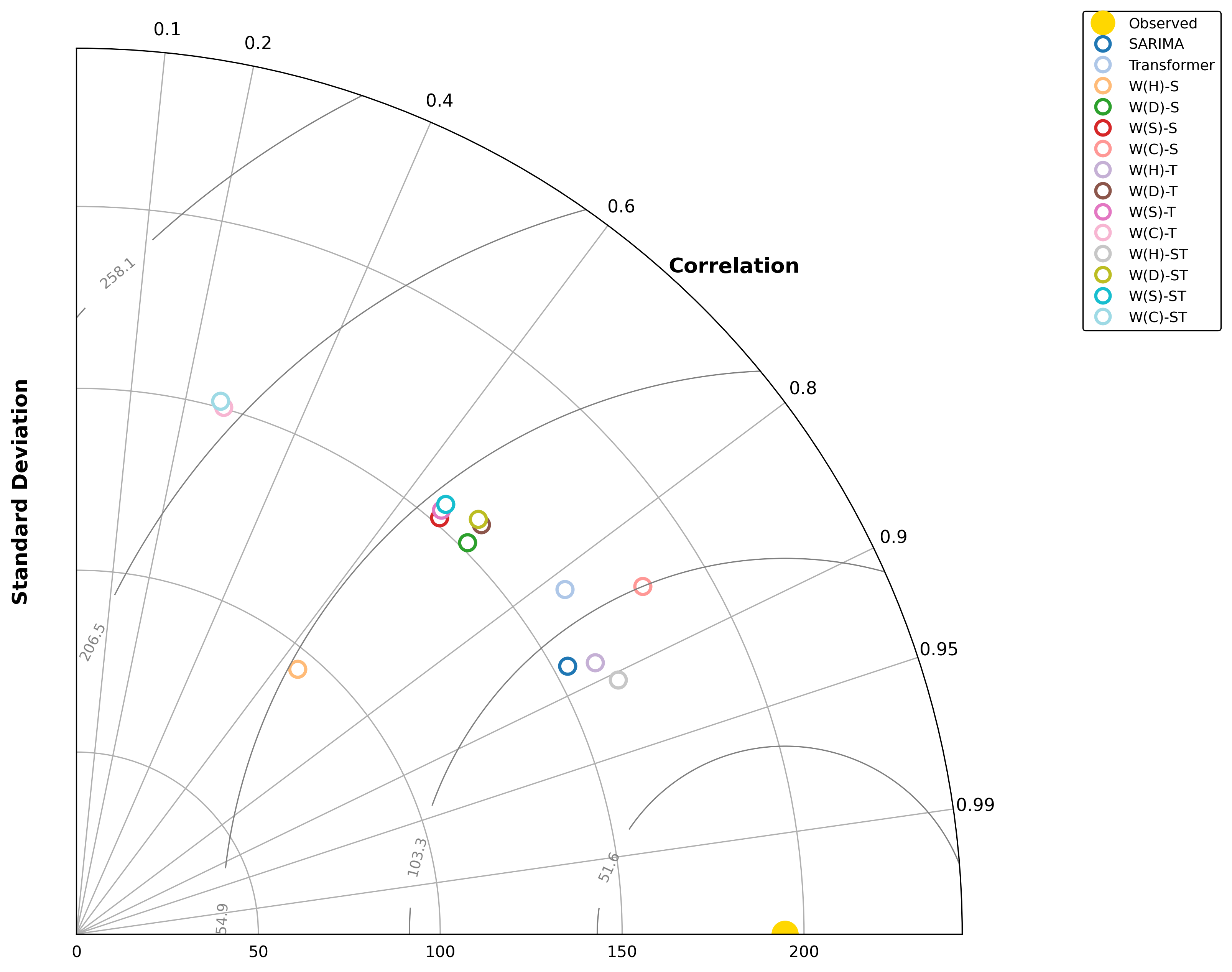}

\includegraphics[width=0.45\textwidth]{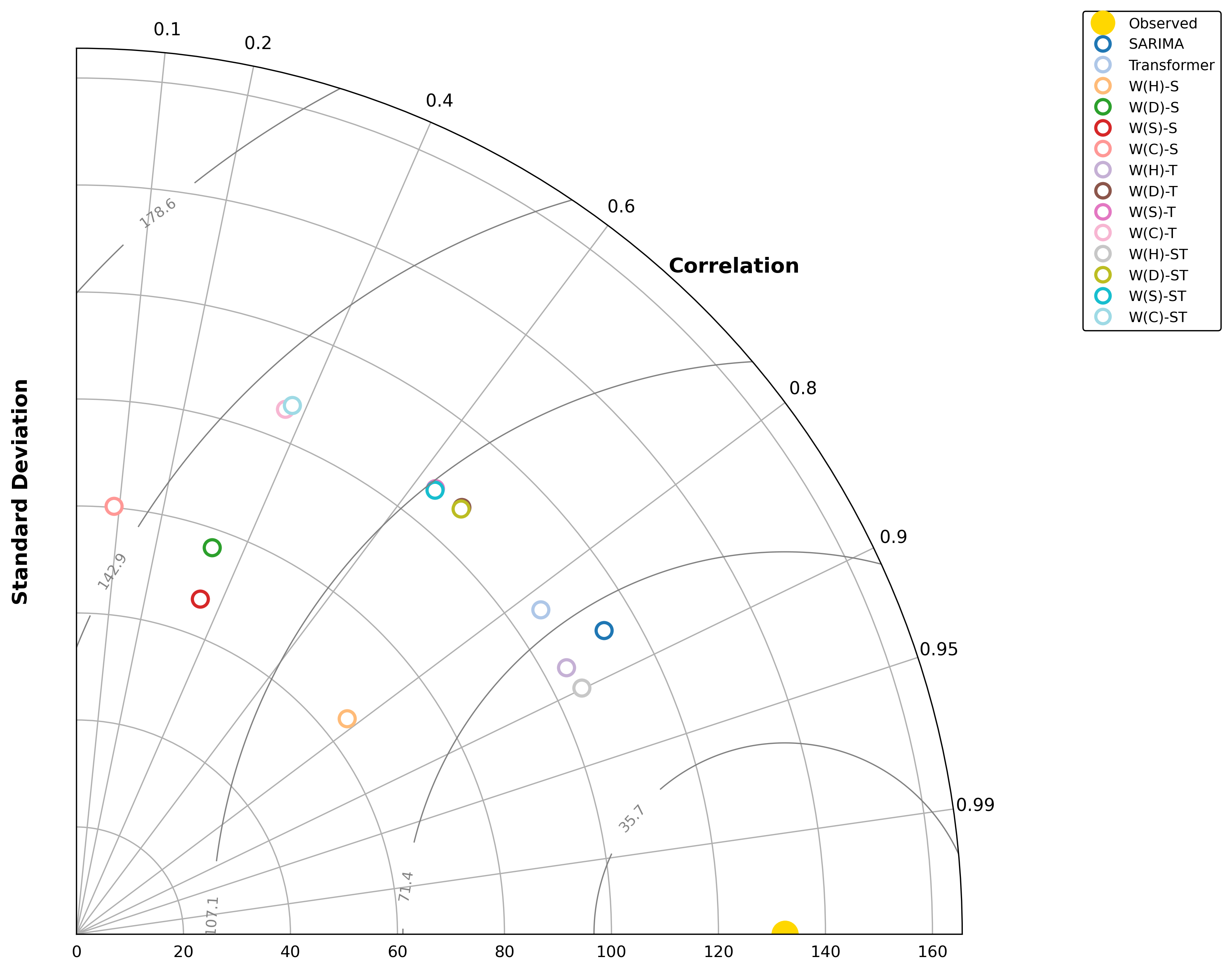}
\includegraphics[width=0.45\textwidth]{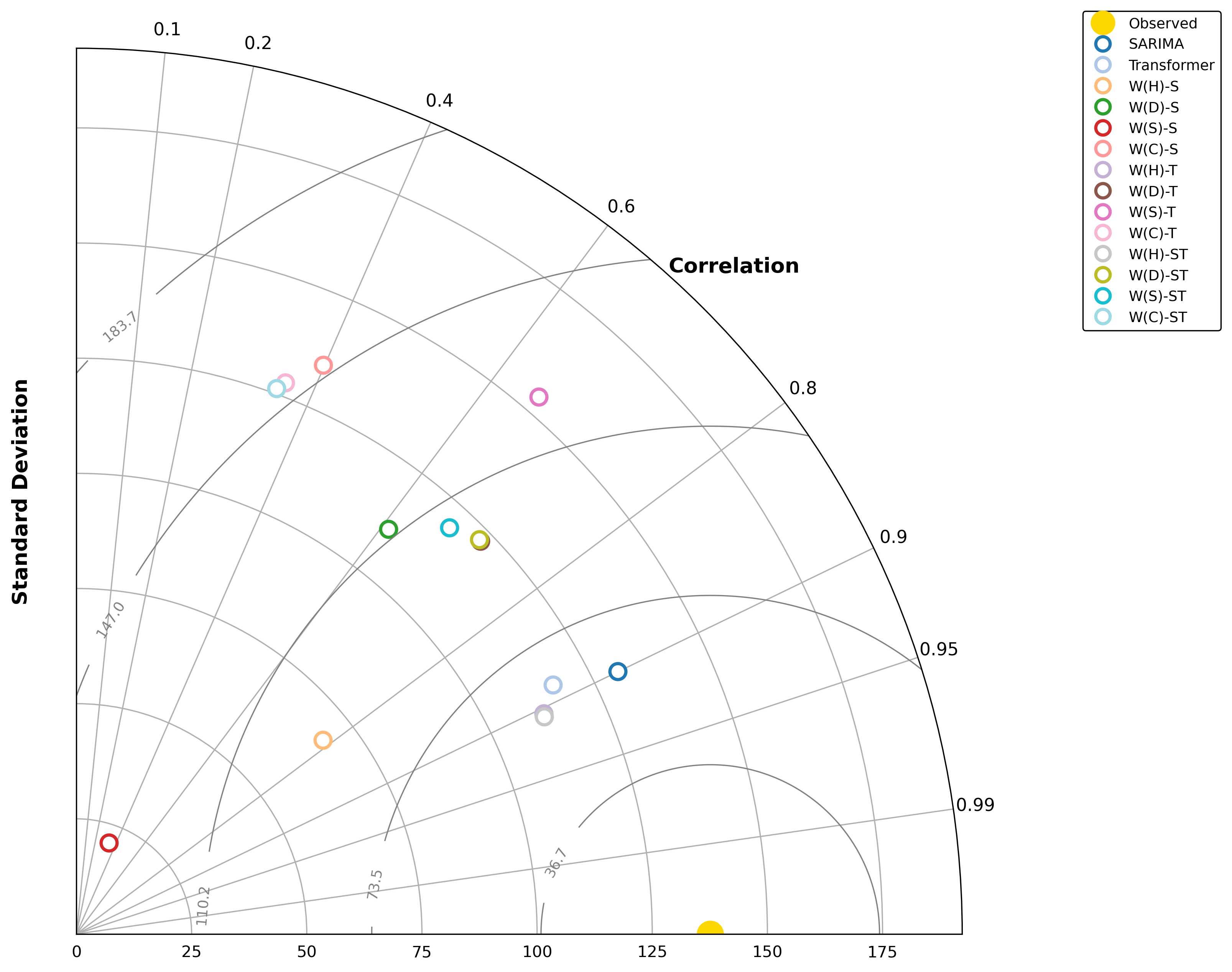}

\includegraphics[width=0.45\textwidth]{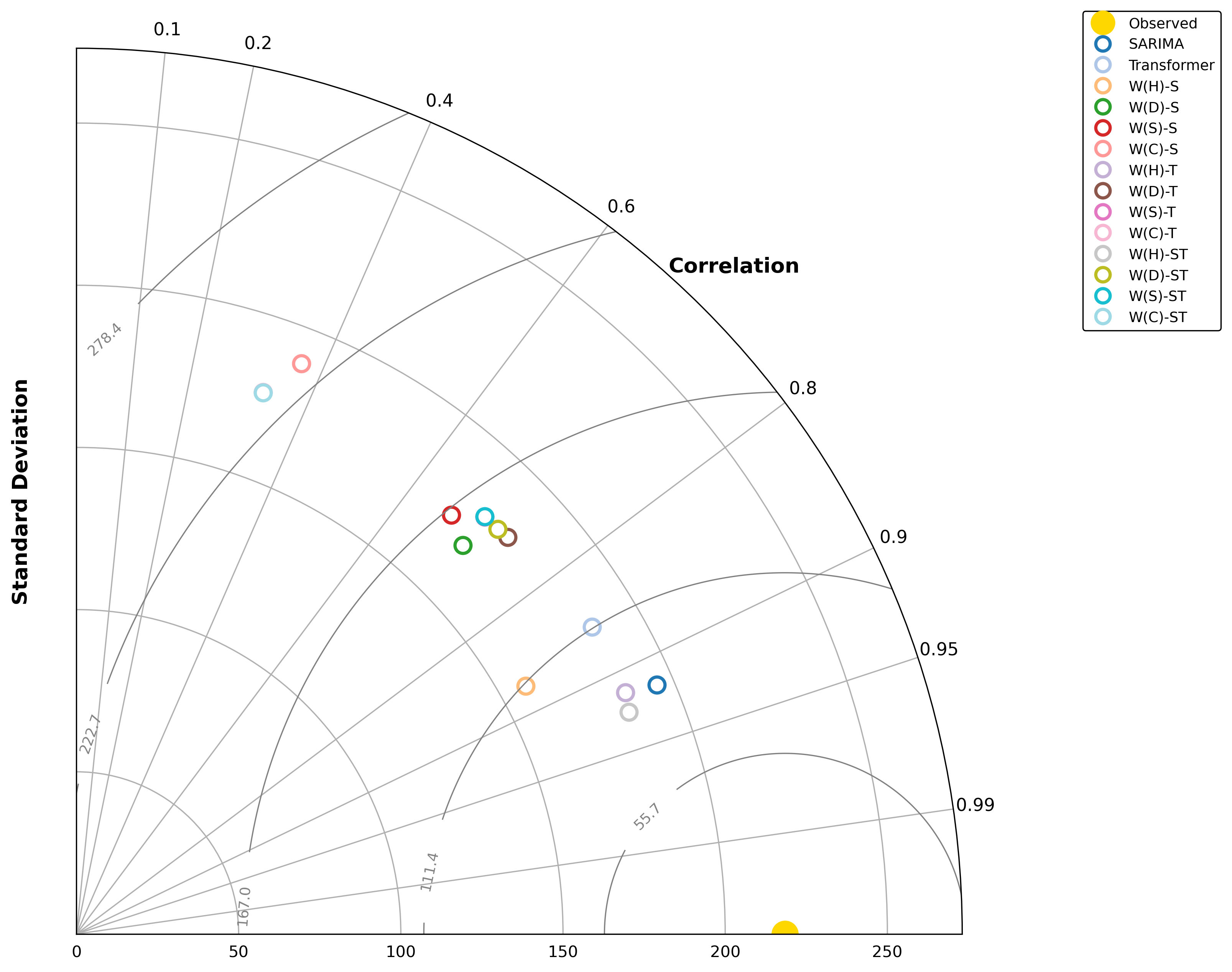}
\caption{Comparison of results across five different Sub-Divisions}
\label{fig:fivefigs}
\end{figure}

The Taylor diagrams (Fig.~\ref{fig:fivefigs}) synthesize three criteria—correlation, standard deviation, and centered RMSE—into a single diagnostic. W(H)-ST and W(D)-ST consistently plot closest to the observation reference point, indicating the best simultaneous alignment in pattern correlation and variance reproduction while minimizing error. By contrast, stand-alone SARIMA often attains respectable correlation but under-represents variability; the stand-alone Transformer better captures nonlinearities but can mis-scale variance without wavelet guidance. Wavelet preprocessing improves both families, but the three-stage hybrid achieves the best overall geometry on the Taylor plots.Tab.~\ref{tab:perfcompare}

\begin{sidewaystable}
\caption{Performance comparison of SARIMA, Transformer and Wavelet-based hybrid models across subdivisions.}
\label{tab:perfcompare}
\begin{tabular*}{\textwidth}{@{\extracolsep\fill}p{.8cm}p{2cm}p{1.2cm}p{1.5cm}p{.8cm}p{.8cm}p{.8cm}p{.8cm}p{.8cm}p{.8cm}p{.8cm}p{.8cm}p{.8cm}p{.8cm}p{.8cm}p{.8cm}}
\toprule%
\textbf{Sub} & \textbf{Metric} & \textbf{SARIMA} & \textbf{Transformer} & \textbf{W(H)-S} & \textbf{W(D)-S} & \textbf{W(S)-S} & \textbf{W(C)-S} & \textbf{W(H)-T} & \textbf{W(D)-T} & \textbf{W(S)-T} & \textbf{W(C)-T} & \textbf{W(H)-ST} & \textbf{W(D)-ST} & \textbf{W(S)-ST} & \textbf{W(C)-ST} \\
\midrule
ARP & Correlation (r) & 0.8288 & 0.7656 & 0.8148 & 0.6479 & 0.6148 & 0.2942 & 0.8637 & 0.675 & 0.6451 & 0.2903 & 0.8821 & 0.6847 & 0.6523 & 0.291 \\
   & Std. Dev (Obs.) & 171.34 & 175.23 & 171.34 & 171.34 & 171.34 & 171.34 & 172.45 & 172.45 & 172.45 & 172.45 & 172.45 & 172.45 & 172.45 & 172.45 \\
   & Std. Dev (Pred.) & 118.91 & 138.53 & 131.62 & 133.13 & 136.28 & 150.89 & 135.38 & 143.20 & 139.82 & 139.74 & 137.30 & 142.04 & 140.43 & 139.93 \\
   & Centered RMSE   & 98.61  & 112.82 & 99.65  & 132.37 & 138.64 & 192.13 & 87.97  & 130.02 & 134.83 & 187.82 & 82.59  & 127.95 & 133.67 & 187.82 \\
\midrule
ASML & Correlation (r) & 0.8776 & 0.8172 & 0.6405 & 0.7067 & 0.6571 & 0.8521 & 0.8860 & 0.7028 & 0.6523 & 0.2687 & 0.9053 & 0.6957 & 0.6514 & 0.2609 \\
     & Std. Dev (Obs.) & 194.79 & 197.95 & 194.79 & 194.79 & 194.79 & 197.30 & 197.13 & 197.13 & 197.13 & 197.13 & 197.13 & 197.13 & 197.13 & 197.13 \\
     & Std. Dev (Pred.) & 153.76 & 164.33 & 94.82  & 152.05 & 151.80 & 182.68 & 160.97 & 158.29 & 153.83 & 150.34 & 164.45 & 158.78 & 155.73 & 151.74 \\
     & Centered RMSE   & 94.96  & 114.12 & 152.56 & 138.56 & 148.76 & 104.28 & 92.44  & 141.61 & 151.53 & 213.38 & 84.88  & 143.26 & 152.05 & 215.11 \\
\midrule
NMMT & Correlation (r) & 0.8996 & 0.8861 & 0.7861 & 0.6106 & 0.3368 & 0.3982 & 0.9046 & 0.7169 & 0.6523 & 0.3542 & 0.9068 & 0.7143 & 0.6761 & 0.3444 \\
     & Std. Dev (Obs.) & 137.60 & 134.45 & 137.60 & 137.60 & 137.60 & 137.60 & 140.50 & 140.50 & 197.13 & 140.50 & 140.50 & 140.50 & 140.50 & 140.50 \\
     & Std. Dev (Pred.) & 130.61 & 116.78 & 68.09  & 110.96 & 20.99  & 134.61 & 112.17 & 122.31 & 153.83 & 127.99 & 111.92 & 122.44 & 119.72 & 126.13 \\
     & Centered RMSE   & 60.48  & 62.36  & 94.02  & 112.25 & 132.02 & 149.35 & 61.71  & 100.30 & 151.53 & 152.91 & 61.23  & 100.78 & 106.43 & 153.11 \\
\midrule
GWB & Correlation (r) & 0.8667 & 0.8197 & 0.7824 & 0.3310 & 0.3461 & 0.0867 & 0.8785 & 0.6701 & 0.6270 & 0.3698 & 0.8989 & 0.6709 & 0.6284 & 0.3779 \\
    & Std. Dev (Obs.) & 132.45 & 131.10 & 132.45 & 132.45 & 132.45 & 132.45 & 135.38 & 135.38 & 135.38 & 135.38 & 135.38 & 135.38 & 135.38 & 135.38 \\
    & Std. Dev (Pred.) & 113.79 & 105.88 & 64.61  & 76.54  & 66.79  & 80.30  & 104.21 & 107.42 & 106.93 & 105.59 & 105.03 & 107.13 & 106.64 & 106.78 \\
    & Centered RMSE   & 66.07  & 75.11  & 91.25  & 129.19 & 126.02 & 148.82 & 66.34  & 101.87 & 107.75 & 137.50 & 61.61  & 101.71 & 107.49 & 137.13 \\
\midrule
SHWBS & Correlation (r) & 0.9189 & 0.8592 & 0.8754 & 0.7051 & 0.6668 & 0.3669 & 0.9153 & 0.7359 & 0.6992 & 0.3258 & 0.9278 & 0.7209 & 0.6991 & 0.3258 \\
      & Std. Dev (Obs.) & 218.46 & 222.06 & 218.46 & 218.46 & 218.46 & 218.46 & 218.62 & 218.62 & 218.62 & 218.62 & 218.62 & 218.62 & 218.62 & 218.62 \\
      & Std. Dev (Pred.) & 194.73 & 184.93 & 158.19 & 168.96 & 173.32 & 189.11 & 184.91 & 180.65 & 179.95 & 176.61 & 183.57 & 180.17 & 180.02 & 176.59 \\
      & Centered RMSE   & 86.38  & 113.76 & 110.65 & 155.62 & 165.13 & 230.59 & 89.36  & 149.33 & 158.63 & 232.01 & 83.80  & 153.18 & 158.66 & 231.99 \\
\botrule
\end{tabular*}
\end{sidewaystable}

That table confirms the Taylor-diagram reading numerically: for example in ARP, correlation improves from 0.829 (SARIMA) to 0.882 [W(H)-ST], predicted SD moves closer to the observed SD (172–175\,mm range), and centered RMSE declines (98.61 $\rightarrow$ 82.59). Similar patterns are seen in ASML and GWB; gains in SHWBS are smaller but still positive.

\subsection{Forecast Verification and Practical Relevance}
\begin{figure}[htbp]
\centering
\includegraphics[width=0.45\textwidth]{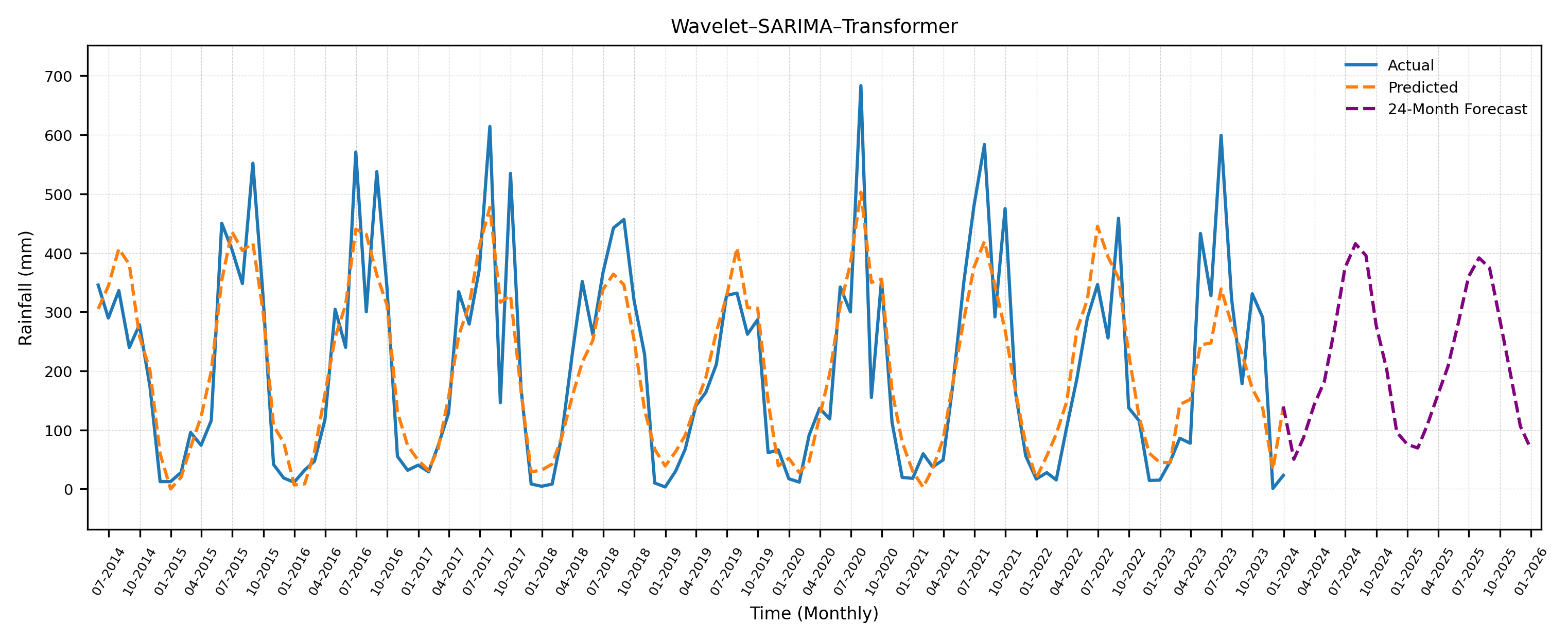}
\includegraphics[width=0.45\textwidth]{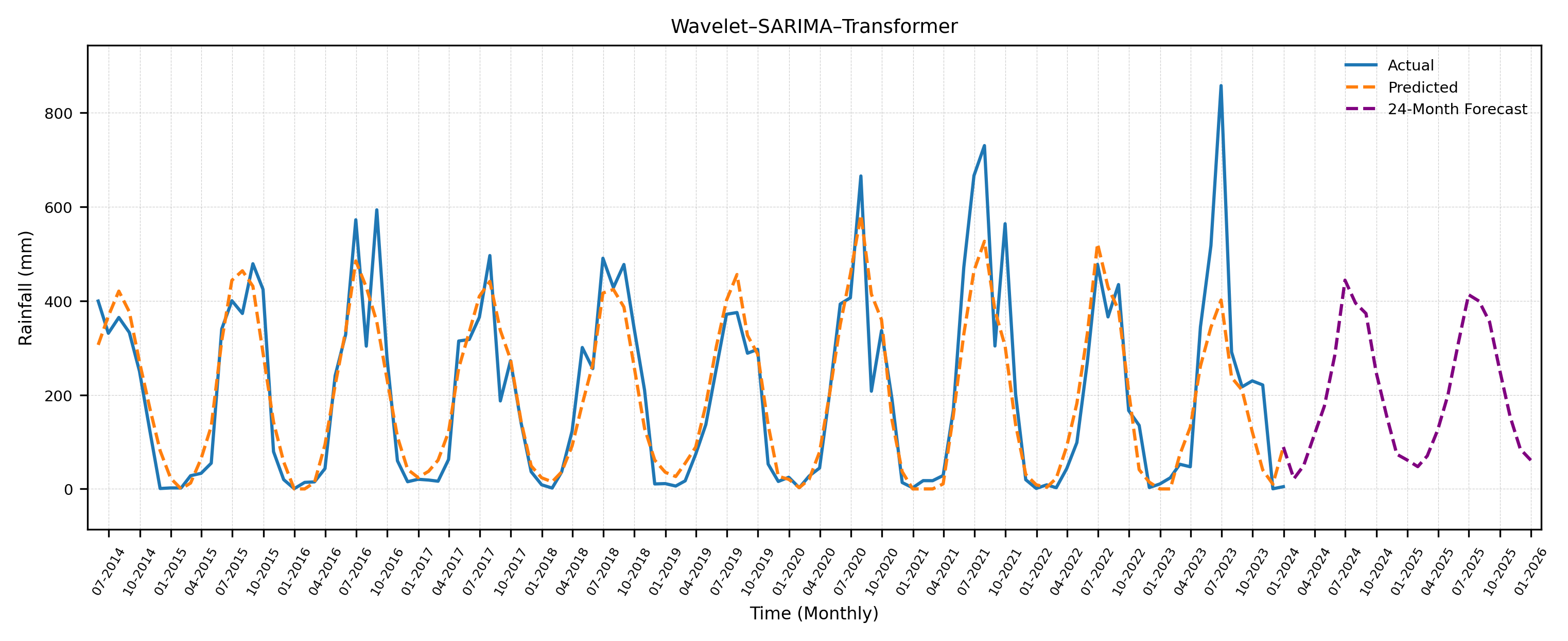}

\includegraphics[width=0.45\textwidth]{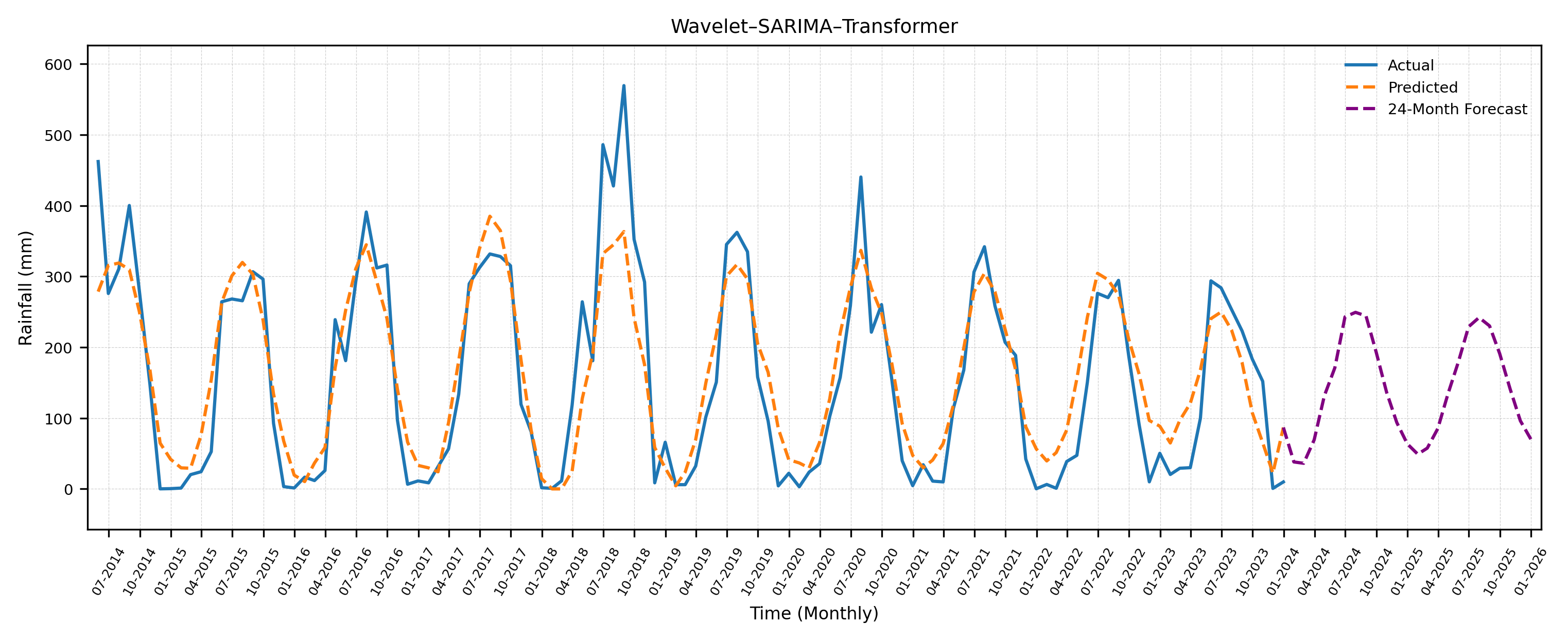}
\includegraphics[width=0.45\textwidth]{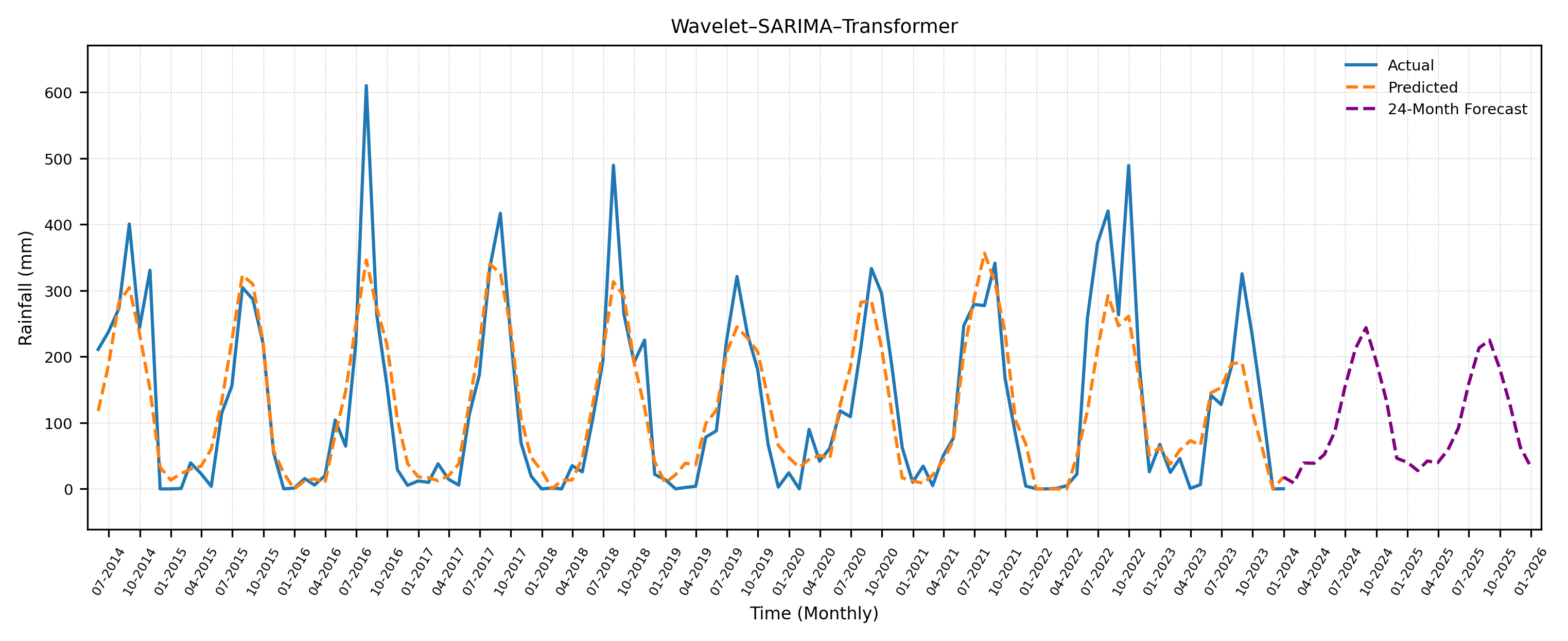}

\includegraphics[width=0.45\textwidth]{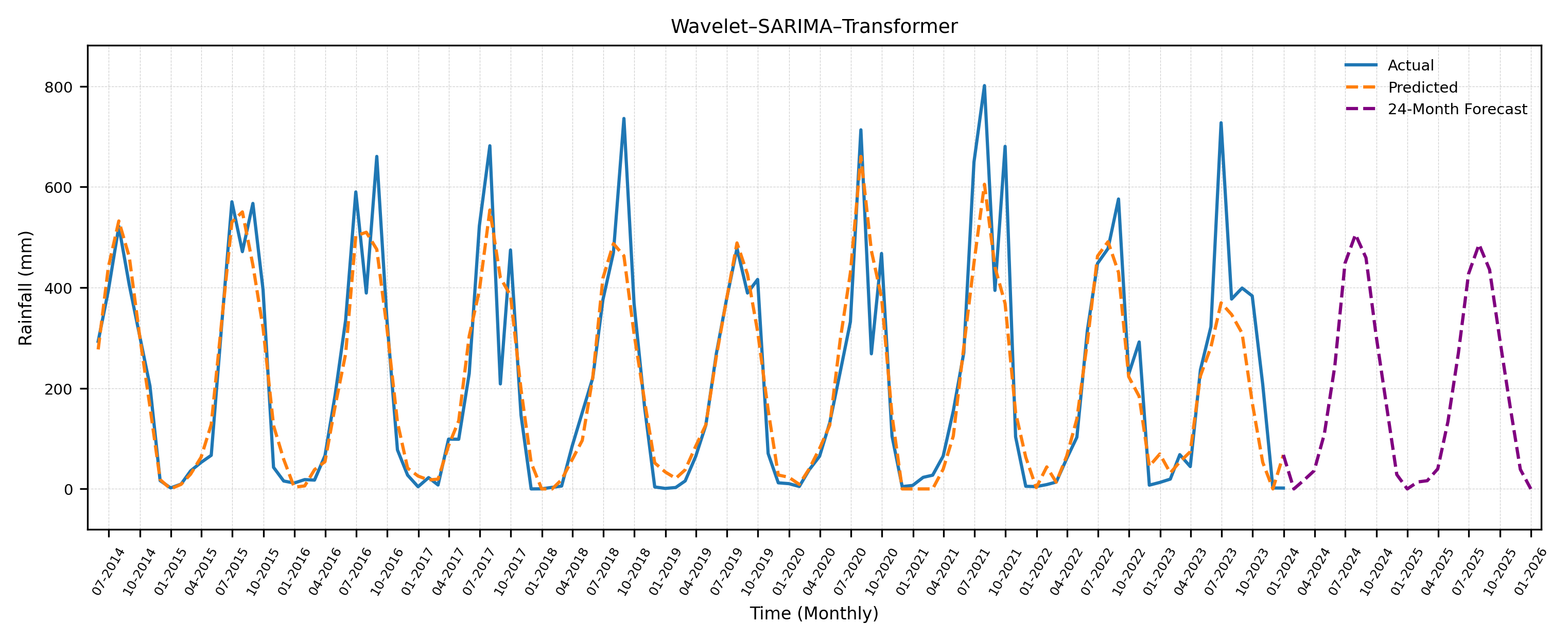}
\caption{Plots of Original vs Predicted}
\label{fig:fivefigsforecast}
\end{figure}

Using W(H)-ST, we generated 24-month ahead forecasts for all five subdivisions. Visual comparisons of original vs.\ predicted series (Fig.~\ref{fig:fivefigsforecast}) show that the hybrid model captures the amplitude and phase of seasonal cycles while preserving transient surges associated with synoptic events. Together with whitened residuals (Table~\ref{tab:ljung}) and superior multimetric performance, this provides convergent evidence of forecasting reliability.

\subsection{What Drives W(H)-ST Superiority?}
Three factors explain the consistent edge:
(i) \textbf{Shift-invariant multiresolution} decomposition (MODWT) aligns scale-specific structures, making seasonal and transient components more learnable;
(ii) \textbf{Model specialization} routes linear seasonal components to SARIMA and nonlinear remnants to a Transformer, reducing misspecification risk;
(iii) \textbf{Reconstruction} (IMODWT) recombines predictions across scales, restoring both large-scale seasonal energy and localized bursts.

\subsection{Sensitivity and Robustness (Summary)}
Performance varies by wavelet family, with Haar and Db4 most frequently yielding top results: Haar excels in regions with abrupt regime shifts; Db4 is preferable under smoother rainfall dynamics. Region-specific variability moderates gains (largest in ARP/ASML; modest but positive in SHWBS), which is typical when baselines already capture a substantial share of seasonal variance.

\subsection{Implications for Hydroclimate Forecasting}
For operational prediction in data-sparse, nonstationary environments, integrating wavelet multiresolution with complementary learners is more effective than any single model class. The W(H)-ST pipeline reduces errors (RMSE/MAE), maintains low bias, increases explained variance, and improves agreement indices—while passing standard adequacy checks. This balanced profile is precisely what is needed for actionable rainfall forecasting and risk planning in the Northeast Indian context.

\section{Conclusion} 
This study proposed and validated a novel hybrid Wavelet–SARIMA–Transformer (W-ST) framework for forecasting monthly rainfall across five meteorological subdivisions of Northeast India. By leveraging the Maximal Overlap Discrete Wavelet Transform (MODWT) for multiresolution decomposition, the model effectively isolated seasonal–linear and nonlinear components, which were then modeled via SARIMA and Transformer networks, respectively.  

A comprehensive evaluation across multiple performance indices demonstrated that the Haar-based hybrid model [W(H)-ST] consistently outperformed stand-alone SARIMA, stand-alone Transformer, and simpler two-stage wavelet hybrids. It achieved lower RMSE and MAE, reduced SMAPE, higher explained variance, stronger agreement indices, and unbiased forecasts (low PBIAS). Residual diagnostic tests confirmed that the models adequately captured serial dependence, with white-noise residuals across all subdivisions. Taylor diagram analyses further reinforced these findings by visually synthesizing correlation, variance fidelity, and RMSE in favor of the proposed hybrid framework.  

The robustness of the W(H)-ST model highlights three methodological strengths: (i) MODWT’s shift-invariant decomposition preserves temporal alignment across scales; (ii) the dual learner design exploits the complementary strengths of SARIMA and Transformer networks; and (iii) multiscale reconstruction ensures both persistent seasonal cycles and transient hydrometeorological bursts are retained.  

From an applied perspective, the framework demonstrates high skill and generalization ability in a region characterized by complex, nonstationary, and extreme rainfall dynamics. Beyond rainfall, the W-ST approach has potential for broader hydroclimatic and environmental time series forecasting, including temperature extremes, drought onset, and flood risk prediction. The methodological advances presented here offer a replicable and scalable template for climate-sensitive regions worldwide, where balancing statistical adequacy with predictive accuracy is critical for operational decision-making and long-term planning.

\bibliographystyle{sn-basic}


\section*{Declarations}
\begin{itemize}
\item Funding:
The authors declare that no funds, grants, or other support were received during the preparation of this manuscript.\\
\item Conflict of interest/Competing interests: 
The authors have no relevant financial or non-financial interests to disclose.\\
\item Author contribution:
Junmoni Saikia, Kuldeep Goswami, and Sarat Chandra Kakaty contributed equally to the study design, data collection, analysis, and manuscript preparation. All authors read and approved the final version of the manuscript.
\end{itemize}

\end{document}